\documentclass[aps,pra,superscriptaddress,twocolumn,amsmath,amsfonts,amssymb,floatfix,nofootinbib]{revtex4}
\usepackage{enumerate}
\usepackage{mathtools,float}
\usepackage{amsmath}
\usepackage{graphicx}
\usepackage{epsfig}
\usepackage{sidecap}
\usepackage{hyperref}
\usepackage[normalem]{ulem}
\usepackage{color}
\usepackage{enumerate}
\usepackage{bm}

\usepackage[utf8]{inputenc}
\usepackage{amsmath}
\usepackage{amsfonts}
\usepackage{amssymb}
\usepackage{graphicx}
\usepackage{enumitem}

\makeatletter

\usepackage[caption=false]{subfig}

\@ifundefined{showcaptionsetup}{}{%
 \PassOptionsToPackage{caption=false}{subfig}}
\usepackage{subfig}
\makeatother

\begin{document}
\title{Interaction of light and semiconductor can generate quantum states
required for solid state quantum computing: Entangled, steered and
other nonclassical states}

\author{Arjun Mukherjee} \thanks{arjun4physics@gmail.com}
\affiliation{Department of Physics, Visva-Bharati, Santiniketan-731 235,
India}

\author{Biswajit Sen} \thanks{bsen75@yahoo.co.in}
\affiliation{Department of Physics, Vidysagar Teachers' Training College,
Midnapore-721 101, India}

\author{Kishore Thapliyal} \thanks{tkishore36@yahoo.com}
\affiliation{Jaypee Institute of Information Technology, A-10, Sector-62, Noida, UP-201309, India}
\affiliation{RCPTM, Joint Laboratory of Optics of Palacky University
and Institute of Physics of Academy of Science of the Czech Republic,
Faculty of Science, Palacky University, 17. listopadu 12, 771 46 Olomouc,
Czech Republic}
   
\author{Swapan Mandal}\thanks{swapanvb@rediffmail.com}
\affiliation{Department of Physics, Visva-Bharati, Santiniketan-731 235,
India}

\author{Anirban Pathak} \thanks{anirban.pathak@gmail.com}
\affiliation{Jaypee Institute of Information Technology, A-10, Sector-62, Noida, UP-201309, India}

\begin{abstract}
Proposals for solid state quantum computing are extremely promising
as they can be used to built room temperature quantum computers. If
such a quantum computer is ever built it would require in-built sources
of nonclassical states required for various quantum information processing
tasks. Possibilities of generation of such nonclassical states are
investigated here for a physical system composed of a monochromatic
light coupled to a two-band semiconductor with direct band gap. The
model Hamiltonian includes both photon-exciton and exciton-exciton
interactions. Time evolution of the relevant bosonic operators are
obtained analytically {{} by using a perturbative technique
that provides operator solution for the coupled Heisenberg's equations
of motion corresponding to the system Hamiltonian. The bosonic operators
are subsequently used to study the possibilities of observing }single
and two mode squeezing and antibunching after interaction in the relevant
modes of light and semiconductor. Further, entanglement between the
exciton and photon modes is reported. Finally, the nonclassical effects
have been studied numerically for the open quantum system scenario.
In this situation, the nonlocal correlations between two modes are
shown to violate EPR steering inequality. The observed nonclassical
features, induced due to exciton-exciton pair interaction, can be
controlled by the phase of input field and the correlations between
two modes are shown to enhance due to nonclassicality in the input
field. 
\end{abstract}
\maketitle

\section{Introduction\label{sec:Introduction}}

During the last few decades, a large number of exciting phenomena
have been observed in the field of quantum optics. These include the generation
of nonclassical states \cite{Glauber,Sudarshan} of radiation field
(e.g., generation of antibunched \cite{anti,anti-mandel}, squeezed
\cite{Loudon_1,Wall}, entangled \cite{Sch-En} and steered \cite{reid-steer}
states of light) and the experimental realization of a set of phenomena
(e.g., laser without inversion \cite{Harris}, enhancement of refractive
index \cite{Scully}, electromagnetically induced transparency \cite{Eit},
absorptionless dispersion \cite{Dowling}, coherent population trapping
\cite{Alzetta}), which were never observed before. The role of squeezed
state in the successful detection of gravitational wave \cite{Abott}
and in the noise free transmissions \cite{Wall}; the advent of quantum
information technology (\cite{nielsen,my book} and references therein)
as well as quantum state engineering \citep{Vogel-SE,Sperling-SE,Miranowicz-SE}
have further enhanced the excitement. Specifically, it may be noted
that quantum supremacy has been strongly established through the realization
of tasks like quantum teleportation \cite{Bennet}, unconditionally
secure quantum cryptography \cite{bb84}, etc. None of these tasks
can be realized in the classical world, and consequently realization
of any of these tasks would require one or more quantum states having
no classical analogue. Such states are referred to as the nonclassical
states and are characterized by the negative values of Glauber-Sudarshan
$P$-function \cite{Glauber,Sudarshan}. 

As nonclassical states can be used to perform tasks that cannot be
done using classical resources, the interest on them is ever increasing.
Possibilities of observing nonclassical features have been investigated
in various physical systems. Specifically, in the recent past nonclassical
properties have been studied in nonlinear optical couplers \cite{kishore-cocoupler,kishore-contra},
Bose-Einstein condensates \cite{sandip-bectwomode,sandip-atommolecule},
Raman \cite{sandip-raman,sandipraman-higherorder} and hyper-Raman
\cite{hyper-Raman} processes, optomechanical and optomechanics-like
systems \cite{optomech}, etc. However, relatively less attention
has been given to the possibilities of generation of nonclassical
states in semiconductors, although that may provide quantum resources
for solid state quantum computing \cite{SolCom,SolCom1,Solcom2,SolCom3}.
Motivated by these facts, in what follows, we study the possibilities
of generation of nonclassical states (e.g., squeezed, antibunched,
entangled, and steering states) through the interaction of monochromatic
light with a two-band semiconductor. 

The possibilities of generation of nonclassical states through the
light-semiconductor interaction have already been investigated in
Refs. \citep{Fox,Rudner,Bascoutas,imam,An+Tran-pla,ba an PRA,BaAn,Ba An with Tinh 2000}.
These investigations were partially motivated by the easy availability
and a few more advantages of the semiconductor over the usual nonlinear
crystals. These studies led to a few interesting results. For example,
by using the optical Kerr effect, the generation of femtosecond pulsed
quadrature squeezed light was reported by Fox et al., in II-VI semiconductor
\cite{Fox}. In addition to this, without going through the cryogenic
cooling and cavity enhancement, the generation of strong squeezing
in a small interaction length was also observed in II-VI semiconductor
\cite{Fox}. The generation of squeezing and entanglement in nuclear
spins were reported in semiconductor quantum dot \cite{Rudner}.
Interestingly, phonon-displaced squeezed number state has been extensively
investigated in polar semiconductors \cite{Bascoutas}. In a recent
experiment, Zeytino\{\textbackslash breve\{g\}\}lu et al., have established
that the direct band gap semiconductor can behave as a nonlinear mirror
and produce broadband squeezing \cite{imam}. Of late, production
of stationary squeezed phonon state is reported in a pulsed optical
excitation of a semiconductor quantum well. On the other hand, nonclassical
properties of excitons of semiconductor and photons of light propagating
through the semiconductor have been studied since last three decades.
For example, the existence of squeezed excitons and/or squeezed radiation
field in such systems has been reported in the nineties \citep{An+Tran-pla,ba an PRA,BaAn,squ-BaAn}.
Almost along the same line, An and Tinh reported that the coupling
of a coherent exciton-photon system may lead to quadrature-amplitude
and number-phase squeezed excitons \cite{ba an polariton added};
they also reported lower-order \cite{Tinh-Nha-An} and higher-order
\cite{Ba An with Tinh 2000} biexciton squeezing for optical exciton-biexciton
conversion; in fact, higher-order squeezing of photon was also reported
in Ref. \cite{An HOS}. 

Keeping in mind the importance of investigating the nonclassical properties
of the radiation field coupled to the semiconductor and the results
of earlier studies, we aim to study a set of hitherto unstudied nonclassical
features which may arise in light-semiconductor interaction. In particular,
in this work, we aim to study the possibilities of observing squeezing
and antibunching of\textcolor{blue}{{} }exciton modes due to interaction
between input coherent light and two-band semiconductor. Further,
in view of the well-known applications of entanglement and steering
in quantum information technology, we would like to investigate the
entanglement and steering between the exciton and photon states, too.

The rest of the paper is organized as follows. In Section \ref{sec:The-Model-Physical},
we describe the physical system of our interest and explicitly provide
a perturbative operator solution of the Heisenberg's equations of
motion corresponding to the Hamiltonian of the physical system of
our interest. Section \ref{sec:Nonclassical-effects} is dedicated
to the nonclassical effects (squeezing, antibunching, and entanglement)
that can be observed in the physical system of our interest. Specifically,
the existence and evolution of these nonclassical features are reported\textcolor{blue}{\sout{.}}
Subsequently, in Section \ref{sec:nonclassicality-under-cw}, the
evolution of the observed nonclassicality is further investigated
in a more general scenario where the field and the semiconductor interact
with their surroundings. In this case, stronger correlation, i.e.,
EPR steering, between exciton and photon modes is also observed. Finally,
the paper is concluded in Section \ref{sec:Conclusion}.

\section{The Model Physical System \label{sec:The-Model-Physical}}

We consider that a monochromatic light, i.e., a single mode of the
electromagnetic field of energy $\omega_{2}$ (in the units $\hbar=1$)
is propagating through a two-band semiconductor with direct band gap,
which is highly excited and allows interband dipole transition. The
interaction of the monochromatic light with this semiconductor would
form a quasi particle, which is referred to as an exciton. As a matter
of fact, the exciton is responsible for transferring the energy to
the electron required for transiting from the valence band to the conduction
band. The exciton is usually viewed as a system composed of interacting
electron-hole pairs. Here, both the electrons and holes are fermionic
in nature. However, the quasiparticle formed because of the formation
of electron-hole pairs are bosonic in nature. Hence, the exciton follows
the bosonic quantization rule, and consequently the Hamiltonian of
the semiconductor coupled to the monochromatic electromagnetic field
can be expressed in the units $\hbar=1$ as \cite{squ-BaAn}
\begin{equation}
\begin{array}{lcl}
H & = & \omega_{1}a^{\dagger}a+\omega_{2}c^{\dagger}c-g\left(a^{\dagger}c+c^{\dagger}a\right)+\chi a^{\dagger2}a^{2},\end{array}\label{1}
\end{equation}
where $a\,(a^{\dagger})$ and $c\,(c^{\dagger})$ are annihilation
(creation) operators for the exciton and photon mode, respectively.
The parameter $g$ corresponds to the photon-exciton interaction constant.
On the other hand, the parameter $\chi$ is the coupling strength
of the exciton-exciton pairs, and hence it is nonlinear in nature.
The bosonic quantization of the exciton have energy $\omega_{1}$
(in the units $\hbar=1$). Some nonclassical features associated with
this Hamiltonian have been studied earlier. For example, by using
the so-called polariton representation of photon and exciton, the
amount of squeezing of light as a function of exciton-exciton coupling
was investigated earlier \cite{squ-BaAn} using secular approximation.
An approximate closed form analytical solutions of this system for
weak nonlinearity (i.e., $\chi\ll1$) was also reported in \cite{Agarwal Puri}.
Further, for strong nonlinearity, a numerical investigation was performed
and the collapse and revival phenomena in the periodic exchange of
energy between atomic oscillator and field were observed. In this
numerical investigation, the relevant bosonic operators were replaced
by their eigenvalues and hence the $c$-numbered differential equations
were formed instead of operator differential equations. This reduced
the computational difficulty and led to an exact result at the cost
of the phase information which is of great significance in the study
of squeezing and entanglement. 

In order to investigate the nonclassical properties of the radiation
field interacting with two band semiconductor, we obtain the Heisenberg's
equations of motion involving operators $a$ and $c$ as
\begin{equation}
\begin{array}{lcl}
\dot{a} & = & -i\left(\omega_{1}a-gc+2\chi a^{\dagger}a^{2}\right),\\
\dot{c} & = & -i\left(\omega_{2}c-ga\right).
\end{array}\label{2}
\end{equation}
These are coupled nonlinear operator differential equations that are
not exactly solvable in closed analytical forms unless $g=0$ and/or
$\chi=0.$ Keeping this in mind, in what follows, we report an approximate
analytical operator solution of these coupled differential equations
(\ref{2}) using Sen-Mandal approach \cite{bsen1,bsen2,bsen3,kishore-cocoupler,kishore-contra}.
This approach is tested in various occasions, and it is established
that it provides better solutions than the solutions obtained by the
well-known short-time approximation technique \cite{Perina-Book,perina-review}
(for a detail discussion on the advantages of Sen-Mandal approach
see \cite{kishore-cocoupler,kishore-contra,hyper-Raman,Zeno}). Here,
we would use Sen-Mandal perturbative technique, to obtain a second
order operator solution. Specially, the solution would be restricted
to the quadratic power of the interaction constants $\chi$ and $g$
considering $gt<1$ and $\chi t<1$. In order to obtain the final
solution, we first follow Sen-Mandal approach to write trial solution
as follows
\begin{equation}
\begin{array}{lcl}
a(t) & = & f_{1}a(0)+f_{2}c(0)+f_{3}a^{\dagger}(0)a^{2}(0)+f_{4}a(0)\\
 & + & f_{5}a^{\dagger}(0)a(0)c(0)+f_{6}c^{\dagger}(0)a^{2}(0)\\
 & + & f_{7}a^{\dagger}(0)a^{2}(0)+f_{8}a^{\dagger2}(0)a^{3}(0),\\
c(t) & = & h_{1}c(0)+h_{2}a(0)+h_{3}c(0)+h_{4}a^{\dagger}(0)a^{2}(0),
\end{array}\label{3}
\end{equation}
where $f_{i}{\rm s}$ and $h_{i}{\rm s}$ are unknown parameters.
It would be relevant to discuss the physical significance of some
of the terms in the above solution (\ref{3}). For example, the term
involving $f_{1}$ is responsible for the free evolution of the exciton
mode $a.$ The term $f_{2}$ exhibits a linear coupling nature between
exciton and photons. The exciton-exciton nonlinear term is responsible
for the appearance of $f_{3}.$ In a similar manner, the appearance
of the remaining $f_{i}{\rm s}$ and $h_{i}{\rm s}$ can also be explained.
Interestingly, the method adopted here does not introduce any restriction
on the terms with higher powers of time $t$ that can be included
in the solution. Specifically, proposed solution may contain terms
with $t^{3}$ and higher powers of $t.$ This is in sharp contrast
with the well-known solutions under short time approximations \cite{Perina-Book,perina-review},
and this is what leads to better results.\textcolor{blue}{{} }In order
to obtain the closed form solution of the evolution of operators $a$
and $c,$ we have to evaluate the unknown parameters $f_{i}{\rm s}$
and $h_{i}{\rm s}$. To do so, we substitute Eq. (\ref{3}) in Eq.
(\ref{2}) and compare the coefficients of same powers of $a(0),$
$c(0),$ etc. to obtain
\begin{equation}
\begin{array}{lcl}
\dot{f_{1}} & = & -i\omega_{1}f_{1},\\
\dot{f_{2}} & = & -i\omega_{1}f_{2}+igh_{1},\\
\dot{f_{3}} & = & -i\omega_{1}f_{3}+2\chi|f_{1}|^{2}f_{1},\\
\dot{f_{4}} & = & -i\omega_{1}f_{4}+igh_{2},\\
\dot{f_{5}} & = & -i\omega_{1}f_{5}-4\chi i|f_{1}|^{2}f_{2},\\
\dot{f_{6}} & = & -i\omega_{1}f_{6}-2\chi if_{1}^{2}f_{2}^{*},\\
\dot{f_{7}} & = & -i\omega_{1}f_{7}+igh_{4}-2\chi i|f_{1}|^{2}f_{3},\\
\dot{f_{8}} & = & -i\omega_{1}f_{8}-4\chi i|f_{1}|^{2}f_{3}-2\chi if_{1}^{2}f_{3}^{*},
\end{array}\label{5}
\end{equation}
and 
\begin{equation}
\begin{array}{lcl}
\dot{h_{1}} & = & -i\omega_{2}h_{1},\\
\dot{h_{2}} & = & -i\omega_{2}h_{2}+igf_{1},\\
\dot{h_{3}} & = & -i\omega_{2}h_{3}+igf_{2},\\
\dot{h_{4}} & = & -i\omega_{2}h_{4}+igf_{3},
\end{array}\label{6}
\end{equation}
respectively.  Finally, we obtain the following analytical expressions
for $f_{i}$s and $h_{i}$s, using the initial conditions $f_{1}(0)=h_{1}(0)=1$
and $f_{i}(0)=h_{i}(0)=0$ for $i\neq1$,
\begin{equation}
\begin{array}{lcl}
f_{1} & = & e^{-i\omega_{1}t},\\
f_{2} & = & \frac{ge^{-i\omega_{1}t}}{\Delta\omega}\left(-1+e^{i\Delta\omega t}\right),\\
f_{3} & = & -2i\chi te^{-i\omega_{1}t},\\
f_{4} & = & \frac{g^{2}e^{-i\omega_{1}t}}{\left(\Delta\omega\right)^{2}}\left(-1+e^{i\Delta\omega t}-it\Delta\omega\right),\\
f_{5} & = & \frac{4g\chi e^{-i\omega_{1}t}}{\left(\Delta\omega\right)^{2}}\left(1-e^{i\Delta\omega t}+it\Delta\omega\right),\\
f_{6} & = & \frac{2g\chi e^{-i\omega_{1}t}}{\left(\Delta\omega\right)^{2}}\left(-1+e^{-i\Delta\omega t}+it\Delta\omega\right),\\
f_{7} & = & f_{8}=-2\chi^{2}t^{2}e^{-i\omega_{1}t},
\end{array}\label{7}
\end{equation}
and 
\begin{equation}
\begin{array}{lcl}
h_{1} & = & e^{-i\omega_{2}t},\\
h_{2} & = & \frac{ge^{-i\omega_{1}t}}{\Delta\omega}\left(-1+e^{i\Delta\omega t}\right)=f_{2},\\
h_{3} & = & \frac{g^{2}e^{-i\omega_{2}t}}{\left(\Delta\omega\right)^{2}}\left(-1+e^{-i\Delta\omega t}+i\Delta\omega t\right),\\
h_{4} & = & \frac{2g\chi e^{-i\omega_{1}t}}{\left(\Delta\omega\right)^{2}}\left(1-e^{i\Delta\omega t}+i\Delta\omega t\right),
\end{array}\label{8}
\end{equation}
where $\Delta\omega=\omega_{1}-\omega_{2}$ is the difference between
the energy of the exciton and photon. Interestingly, the absence of
exciton-exciton interaction causes $f_{3}=f_{5}=f_{6}=f_{7}=f_{8}=h_{4}=0.$
The consistency of the obtained solution is further checked by confirming
that the obtained solution satisfies equal time commutation relation
$[a(t),a^{\dagger}(t)]$ = $[c(t),c^{\dagger}(t)]$ = $1$ and $[a(t),c^{\dagger}(t)]=[a(t),c(t)]$
=$0$. The obtained solution may now be used to calculate the expectation
values of various operators that are required for the present investigation
on the nonclassical behavior of the exciton and field modes considering
that the initial state is a composite coherent state 
\begin{equation}
|\psi(0)\rangle=|\alpha\rangle|\beta\rangle.\label{9}
\end{equation}
This composite coherent state can be viewed as a product of two coherent
states $|\alpha\rangle$ and $|\beta\rangle$, such that $a\left(0\right)|\alpha\rangle=\alpha|\alpha\rangle$
and $c\left(0\right)|\beta\rangle=\beta|\beta\rangle.$ For example,
for this initial state, we can obtain the expectation values of the
number operators $\left\langle N_{a}(t)\right\rangle =\left\langle a^{\dagger}(t)a(t)\right\rangle $
and $\left\langle N_{c}(t)\right\rangle =\left\langle c^{\dagger}(t)c(t)\right\rangle $
as follows
\begin{equation}
\begin{array}{lcl}
\left\langle N_{a}(t)\right\rangle  & = & \left|\alpha\right|^{2}+\left|f_{2}\right|^{2}\left(\left|\beta\right|^{2}-\left|\alpha\right|^{2}\right)\\
 & + & \left\{ f_{1}^{*}f_{2}\alpha^{*}\beta-h_{1}^{*}h_{4}\left|\alpha\right|^{2}\alpha\beta^{*}+{\rm c.c.}\right\} 
\end{array}
\end{equation}
and
\begin{equation}
\begin{array}{lcl}
\left\langle N_{c}(t)\right\rangle  & = & \left|\beta\right|^{2}+\left|h_{2}\right|^{2}\left(\left|\alpha\right|^{2}-\left|\beta\right|^{2}\right)+\left\{ \left(h_{1}^{*}h_{2}\beta^{*}\alpha\right.\right.\\
 & + & \left.\left.h_{1}^{*}h_{4}\left|\alpha\right|^{2}\alpha\beta^{*}\right)+{\rm c.c.}\right\} ,
\end{array}
\end{equation}
where c.c. stands for complex conjugate. In what follows, we consider
$\alpha$ as real, but $\beta=\left|\beta\right|e^{-i\phi}$ as complex,
where $\phi$ is the phase angle of the input coherent state corresponding
to the photon mode. Further, $|\alpha|^{2}=\alpha^{2}$ would represent
the initial number of exciton of the system, whereas $|\beta|^{2}$
is the average photon number before the interaction.

\section{Nonclassical effects \label{sec:Nonclassical-effects}}

In the previous section, we have already obtained the analytical solution
for the dynamics of the relevant bosonic operators. In the present
section, we will use the obtained solution for investigating the nonclassical
features present in the physical system of our interest. To begin
with, in the following subsection, we investigate the possibility
of observing quadrature squeezing.

\subsection{Quadrature Squeezing}

\begin{figure}
\subfloat[]{\includegraphics[scale=0.45]{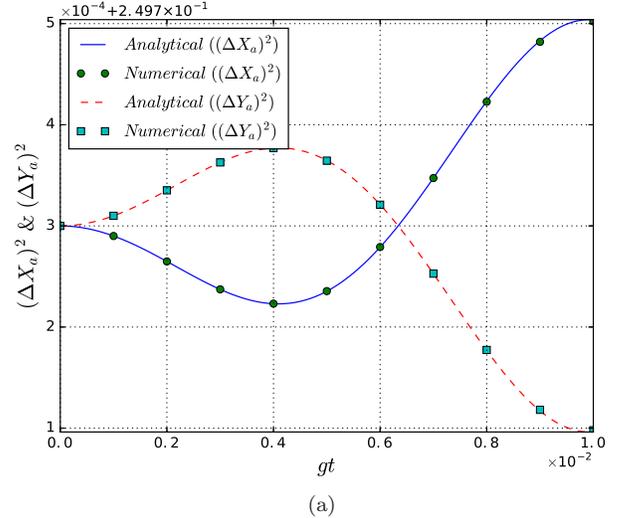}

}

\subfloat[]{\includegraphics[scale=0.45]{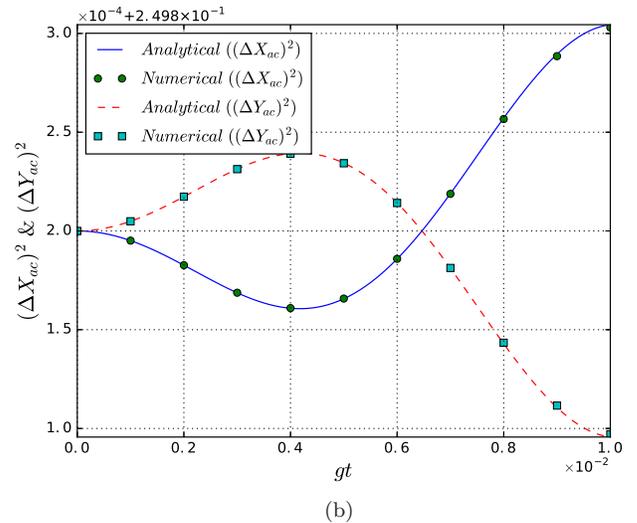} 

} 

\caption{\label{fig.squeezing}(Color online) Variation of quadrature squeezing
with rescaled time $gt$ using parameters (mentioned in text) that
are consistent with the real systems (for CdS) \cite{ba an PRA},
and $\alpha=2.0,$ $\beta=1.0$, for (a) exciton mode $a$ and (b)
the compound mode $ac$. All the quantities shown here and the rest
of the figures are dimensionless. }
\end{figure}

The dimensionless quadrature operators are defined in terms of the
usual annihilation and creation operators as follows
\begin{equation}
\begin{array}{lcl}
X_{k} & = & \frac{1}{2}\left[k(t)+k^{\dagger}(t)\right],\\
Y_{k} & = & \frac{1}{2i}\left[k(t)-k^{\dagger}(t)\right],
\end{array}\label{12}
\end{equation}
where $k$ ($k^{\dagger}$) is the annihilation (creation) operator
of a particular mode. If the variances $\left(\Delta X_{k}\right)^{2}$
or $\left(\Delta Y_{k}\right)^{2}$ goes below $\frac{1}{4},$ the
corresponding quadrature is said to be squeezed. It is clear that
the simultaneous squeezing of both the quadrature is prohibited by
Heisenberg's uncertainty principle. Using Eqs. (\ref{3}), (\ref{7})-(\ref{9}),
and (\ref{12}), we obtain the detailed analytical expressions for
the variances involving the exciton mode $a$ and the photon mode
$c$ as follows
\begin{equation}
\begin{array}{lcl}
\left[\begin{array}{c}
\left(\Delta X_{a}\right)^{2}\\
\left(\Delta Y_{a}\right)^{2}
\end{array}\right] & = & \frac{1}{4}\left[1+2\left|f_{3}\right|^{2}\left|\alpha\right|^{4}+\left\{ \left(f_{1}^{*}f_{5}+2f_{1}f_{6}^{*}\right)\alpha^{*}\beta\right.\right.\\
 & + & \left.{\rm c.c.}\right\} \pm\left\{ \left(f_{1}f_{3}+f_{1}f_{7}\right)\alpha^{2}+f_{1}f_{5}\alpha\beta\right.\\
 & + & \left.2f_{1}f_{8}\left|\alpha\right|^{2}\alpha^{2}+{\rm c.c.}\right\} ,
\end{array}\label{13}
\end{equation}
and
\begin{equation}
\begin{array}{lcl}
\left[\begin{array}{c}
\left(\Delta X_{c}\right)^{2}\\
\left(\Delta Y_{c}\right)^{2}
\end{array}\right] & = & \frac{1}{4}\end{array}.\label{14}
\end{equation}
It is clear from Eq. (\ref{14}) that no squeezing is observed in
photon mode $c$. However, it is possible to observe the squeezing
effects of the exciton mode $a$. Interestingly, for $\chi=0,$ the
squeezing in exciton mode is found to disappear. It substantiates
that the effective nonlinearity induced by the exciton-exciton interaction
is indeed responsible for the squeezing phenomenon.\textcolor{red}{{}
}The analytical expression (\ref{13}) is now at our disposal to investigate
the possibility of observing squeezing of the exciton mode and the
evolution of it as a function of the exciton number, photon number
and/or the phase of the field mode. However, we restrict ourselves
to the study of evolution of $\left(\Delta X_{a}\right)^{2}$ and
$\left(\Delta Y_{a}\right)^{2}$ with the rescaled time $gt$ only
in Fig. \ref{fig.squeezing} (a). The figure clearly illustrates the
existence of squeezing. In addition to the perturbative analytical
expressions for the variance, we have also obtained the exact numerical
values for the variance using QuTiP 3.1.0 \citep{qutip1,qutip2} by
solving time dependent Schr{\"o}dinger equation using the matrix
forms of operators and initial state. The results obtained through
the numerical analysis are represented by the squares and circles.
It is clear from Fig. \ref{fig.squeezing} (a) (and a set of other
figures included in this paper, where both numerical and analytical
investigations have been conducted) that the numerical results exactly
coincide with the corresponding analytical results obtained here. 

In Fig. \ref{fig.squeezing} and in the rest of the paper we have
chosen the parameters in analogy with the earlier works of An and
others \cite{ba an PRA,Tinh-Nha-An,ba an polariton added,An HOS,An+Tran-pla}.
The parameters are selected in such a way that they remain consistent
with a real physical system. Specifically, we have considered that
the semiconductor of our system is CdS and for all the plots, we have
used following parameters (which correspond to CdS): $\omega_{1}=25.277g$,
$\omega_{2}$ =$24.013g$, and $\chi=5.304g$. Here, it may be apt
to note that in Ref. \cite{ba an PRA}, these values of the parameters
were obtained using the relations $E_{e}(k)=E_{g}+k^{2}/(2m_{e})$
and $E_{h}(k)=k^{2}/(2m_{h})$, where $E_{g},\,E_{e}(k),$ and $E_{h}(k)$
are the band gap energy, energy of the electron, and energy of the
hole, respectively. Also, $m_{e}(m_{h})$ is the effective mass of
the electron (hole). 

Through Fig. \ref{fig.squeezing} (a) and Eq. (\ref{14}), we can
recognize that squeezing can be observed in exciton mode, but not
in the photon mode. Therefore, it will be interesting to investigate
the possibility of intermodal squeezing. In order to do so, we use
the following quadrature operators involving the exciton and photon
mode \cite{Loudon_1}:
\begin{equation}
\begin{array}{lcl}
X_{ac} & = & \frac{1}{2\sqrt{2}}\left[a(t)+a^{\dagger}(t)+c(t)+c^{\dagger}(t)\right],\\
Y_{ac} & = & \frac{1}{2\sqrt{2}i}\left[a(t)-a^{\dagger}(t)+c(t)-c^{\dagger}(t)\right],
\end{array}\label{15}
\end{equation}
and the following criterion of intermodal squeezing $\left(\Delta X_{ac}\right)^{2}<\frac{1}{4}$
or $\left(\Delta Y_{ac}\right)^{2}<\frac{1}{4}$. 

A bit of calculation using Eqs. (\ref{3}), (\ref{7})-(\ref{9}),
and (\ref{15}) yields analytical expression for the variance of the
coupled mode quadratures as
\begin{equation}
\begin{array}{lcl}
\left[\begin{array}{c}
\left(\Delta X_{ac}\right)^{2}\\
\left(\Delta Y_{ac}\right)^{2}
\end{array}\right] & = & \frac{1}{4}\left[1+\left|f_{3}\right|^{2}\left|\alpha\right|^{4}+\frac{1}{2}\left\{ f_{1}^{*}h_{2}+h_{1}f_{2}^{*}\right.\right.\\
 & + & \left(f_{1}^{*}f_{5}+2f_{1}f_{6}^{*}\right)\alpha^{*}\beta+\left(2f_{1}h_{4}^{*}+h_{1}f_{5}^{*}\right.\\
 & + & \left.2f_{3}h_{2}^{*}\right)\left|\alpha\right|^{2}\pm\left(f_{1}f_{3}+f_{1}f_{7}+f_{1}h_{4}+h_{1}f_{6}\right.\\
 & + & \left.\left.\left.h_{2}f_{3}\right)\alpha^{2}+f_{1}f_{5}\alpha\beta+3f_{3}^{2}\left|\alpha\right|^{2}\alpha^{2}+{\rm c.c.}\right\} \right].
\end{array}\label{16}
\end{equation}
The right-hand side of Eq. (\ref{16}) is plotted in Fig. \ref{fig.squeezing}
(b), which clearly shows the existence of intermodal squeezing in
both the quadratures of the coupled mode (of course not simultaneously).
The plot also shows that our perturbative analytical results considerably
match with the corresponding exact numerical results.

\subsection{Antibunching}

\begin{figure}
\subfloat[]{\includegraphics[scale=0.45]{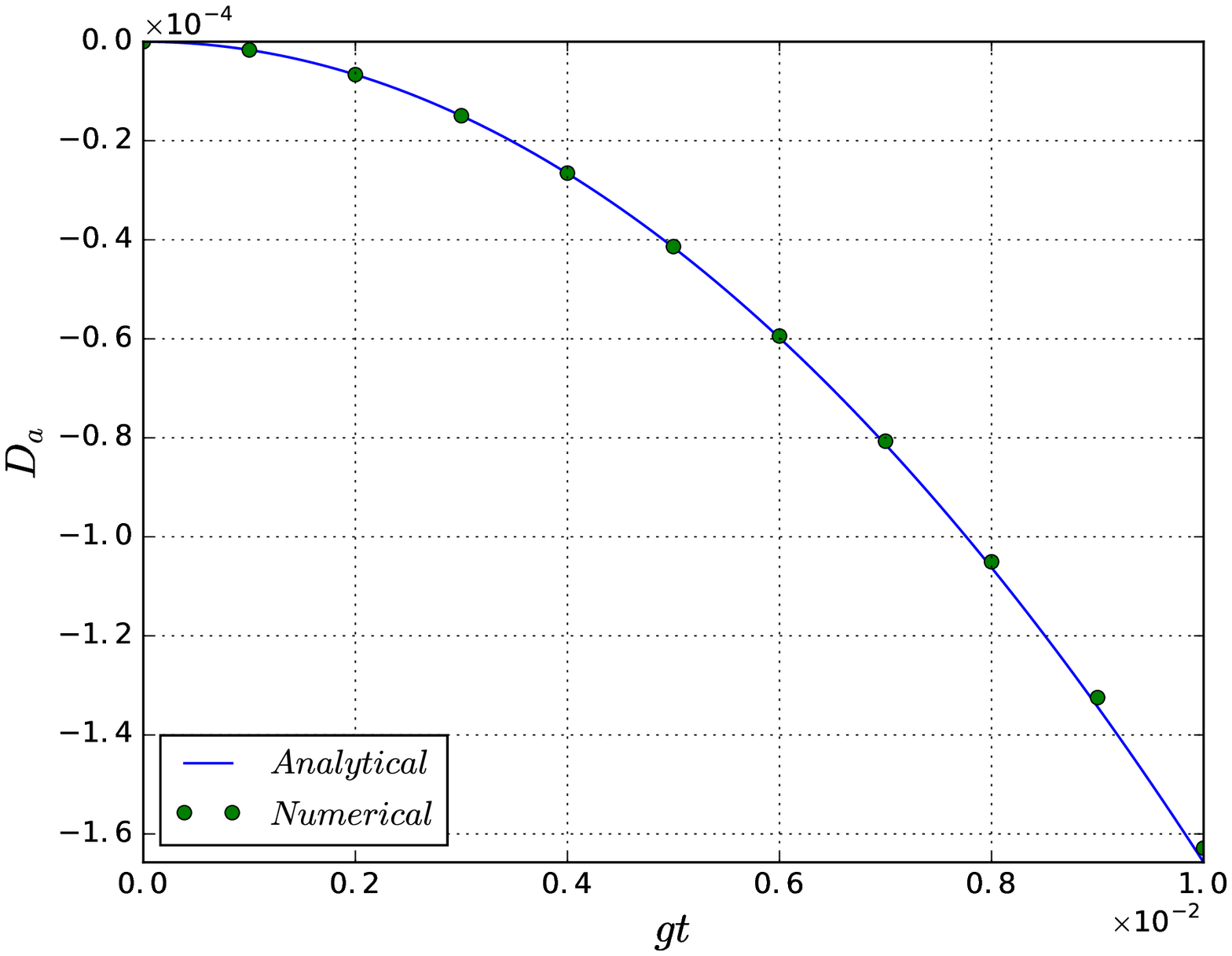}

}

\subfloat[]{\includegraphics[scale=0.45]{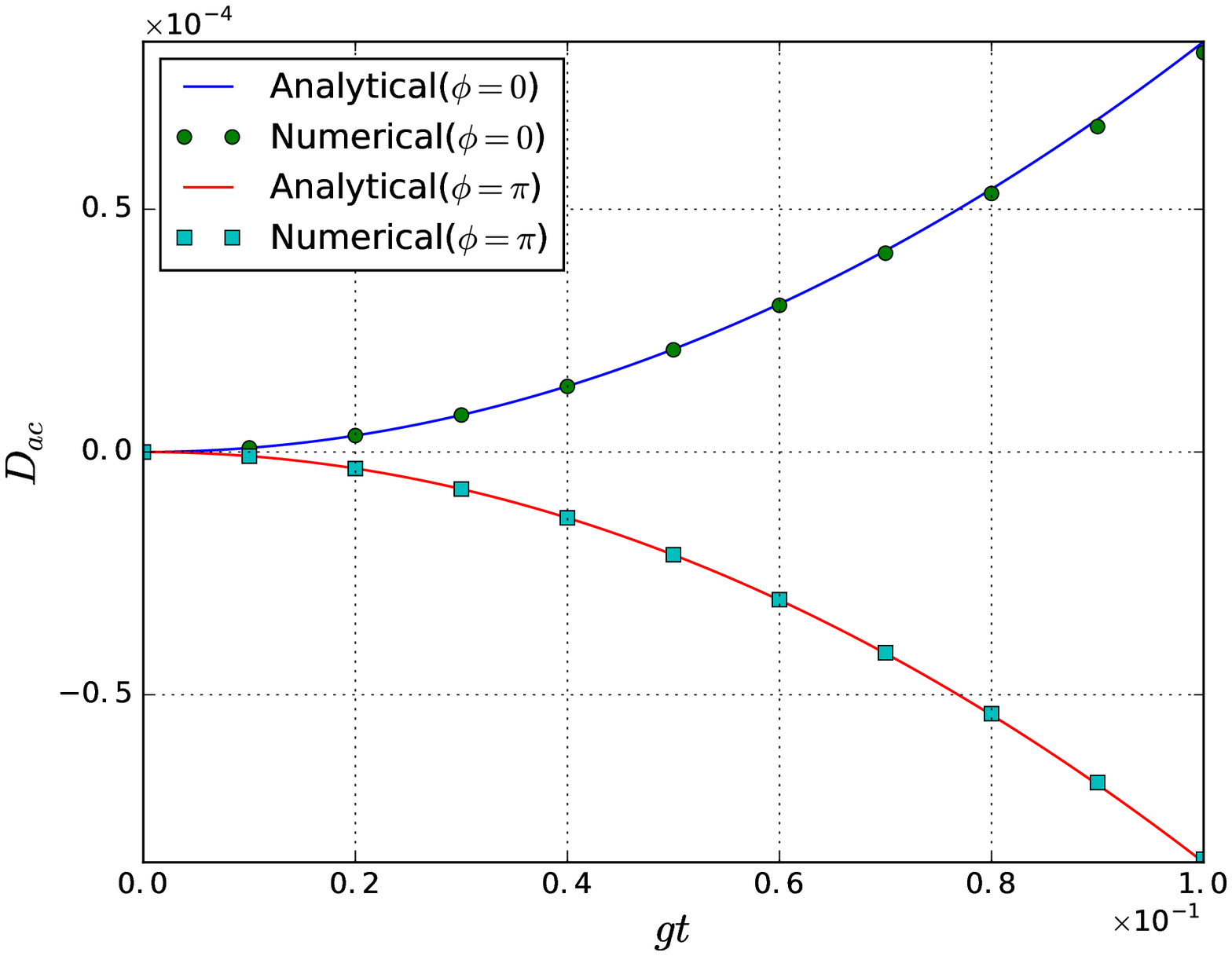}

} 

\caption{\label{figantibunching} (Color online) Variation of antibunching
with $gt$ using the same set of values of parameters as in Fig. \ref{fig.squeezing}
for (a) exciton mode $a$ and (b) the compound mode $ac$ (for two
values of the phase angle $\phi=0$ and $\pi$).}
\end{figure}

One of the most intensively studied nonclassical phenomena is antibunching
of bosons. The photons from an incandescent light source come in a
bunch and show bunching. However, there are many physical systems,
where antibunching of bosons can be observed \cite{anti,anti-mandel,kishore-cocoupler,kishore-contra,hyper-Raman}.
The theoretical investigation on the antibunching of bosons are usually
performed using the so-called second order correlation function at
zero time delay, which is defined as
\begin{equation}
\begin{array}{lcl}
g_{i}^{(2)}(0) & =\frac{\left\langle i^{\dagger}(t)i^{\dagger}(t)i(t)i(t)\right\rangle }{\left\langle i^{\dagger}(t)i(t)\right\rangle \left\langle i^{\dagger}(t)i(t)\right\rangle }= & 1+\frac{D_{i}}{\left\langle N_{i}(t)\right\rangle ^{2}},\end{array}\label{17}
\end{equation}
where 
\begin{equation}
D_{i}=\left(\Delta N_{i}(t)\right)^{2}-\left\langle N_{i}(t)\right\rangle ,\label{18}
\end{equation}
and $i\in\{a,c\}$. The condition for the single mode antibunching
in $i^{\mathrm{th}}$ mode is $g_{i}^{(2)}(0)<1$, which is equivalent
to $D_{i}<0$, and generally it implies antibunching as well as sub-Poissonian
boson (photon/exciton) statistics. In fact, $g_{i}^{(2)}(0)$ is more
associated with sub-Poissonian boson statistics and ensures the presence
of antibunching at least for some time scales (\citep{kishore-cocoupler}
and references therein). Therefore, we have followed this convention
here, and have used $D_{i}<0$ as the criterion of antibunching. Using
Eqs. (\ref{3}), (\ref{7})-(\ref{9}), and (\ref{18}), we can easily
obtain an analytical expression for $D_{i}$ as follows
\begin{equation}
\begin{array}{lcl}
D_{a} & = & \left\{ 2\left(f_{1}^{*}f_{5}+f_{1}f_{6}^{*}+2f_{2}f_{3}^{*}+f_{1}^{*2}f_{2}f_{3}\right)\right.\\
 & \times & \left.\left|\alpha\right|^{2}\alpha^{*}\beta+{\rm c.c.}\right\} 
\end{array}\label{19}
\end{equation}
and

\begin{equation}
\begin{array}{lcl}
D_{c} & = & \left|f_{2}\right|^{2}\left|\beta\right|^{2}.\end{array}\label{20}
\end{equation}
It is clear from the analytical expression of $D_{c}$\textcolor{red}{{}
}that the photon mode always remains bunched (at least in the domain
of the validity of the perturbative solution) after the interaction,
which also establishes that it exhibits super-Possonian photon statistics.
This is so because the right-hand side of Eq. (\ref{20}) is always
positive. However, the exciton mode may be antibunched. In order to
study the time evolution of the observed nonclassical property, we
plot $D_{a}$ as a function of dimensionless interaction time $gt$
in Fig. \ref{figantibunching} (a). Interestingly, for the chosen
set of values of parameters, the antibunching in exciton mode is exhibited.
The depth of the antibunching witnessing parameter $D_{a}$ is observed
to increase with increasing $gt.$ The monotonic nature of $D_{a}$
could be attributed to the truncated nature of the solutions for the
bosonic operators. Further, the condition of intermodal antibunching
can be expressed as
\begin{equation}
\begin{array}{lcl}
D_{ac} & = & \left\langle a^{\dagger}(t)c^{\dagger}(t)c(t)a(t)\right\rangle -\left\langle a^{\dagger}(t)a(t)\right\rangle \left\langle c^{\dagger}(t)c(t)\right\rangle .\end{array}\label{22}
\end{equation}
Here, $D_{ac}<0$, $D_{ac}>0$, and $D_{ac}=0$ correspond to intermodal
antibunching, bunching and unbunched two mode state, respectively.
The analytical expression for the intermodal antibunching witnessing
parameter $D_{ac}$ can be computed as 
\begin{equation}
\begin{array}{lcl}
D_{ac} & = & \left[\left(f_{1}f_{6}^{*}+f_{1}f_{3}^{*}h_{1}h_{2}^{*}\right)\left|\alpha\right|^{2}\alpha^{*}\beta+{\rm c.c.}\right]\end{array}.\label{23}
\end{equation}
In order to get the flavor of the analytical expression, we plot the
right-hand side of Eq. (\ref{23}) for two values of the phase angle
(i.e., for $\phi=0$ and $\phi=\pi$) of the input radiation field
in the coherent state. The figure establishes that the occurrence
of nonclassciality (in this case, intermodal antibunching) can be
controlled by the phase of the input radiation field. Specifically,
intermodal antibunching is observed for $\phi=\pi$, while we could
not find it for $\phi=0$. Also, the analytical results obtained here
match exactly with the numerical results which further establishes
the validity of our analytical solution.

\begin{figure}
\subfloat[]{\begin{centering}
\includegraphics[scale=0.45]{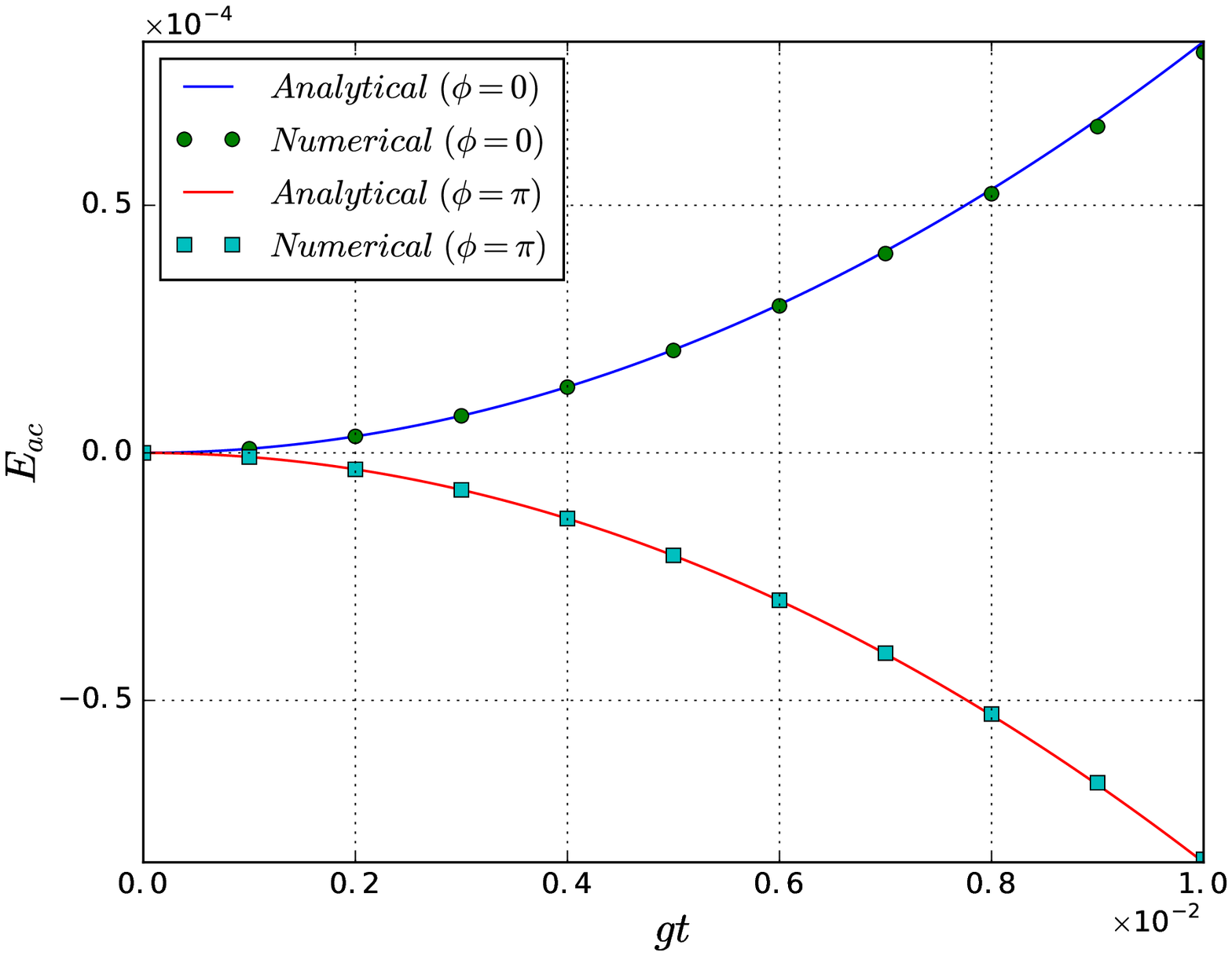} 
\par\end{centering}
}

\subfloat[]{\begin{centering}
\includegraphics[scale=0.45]{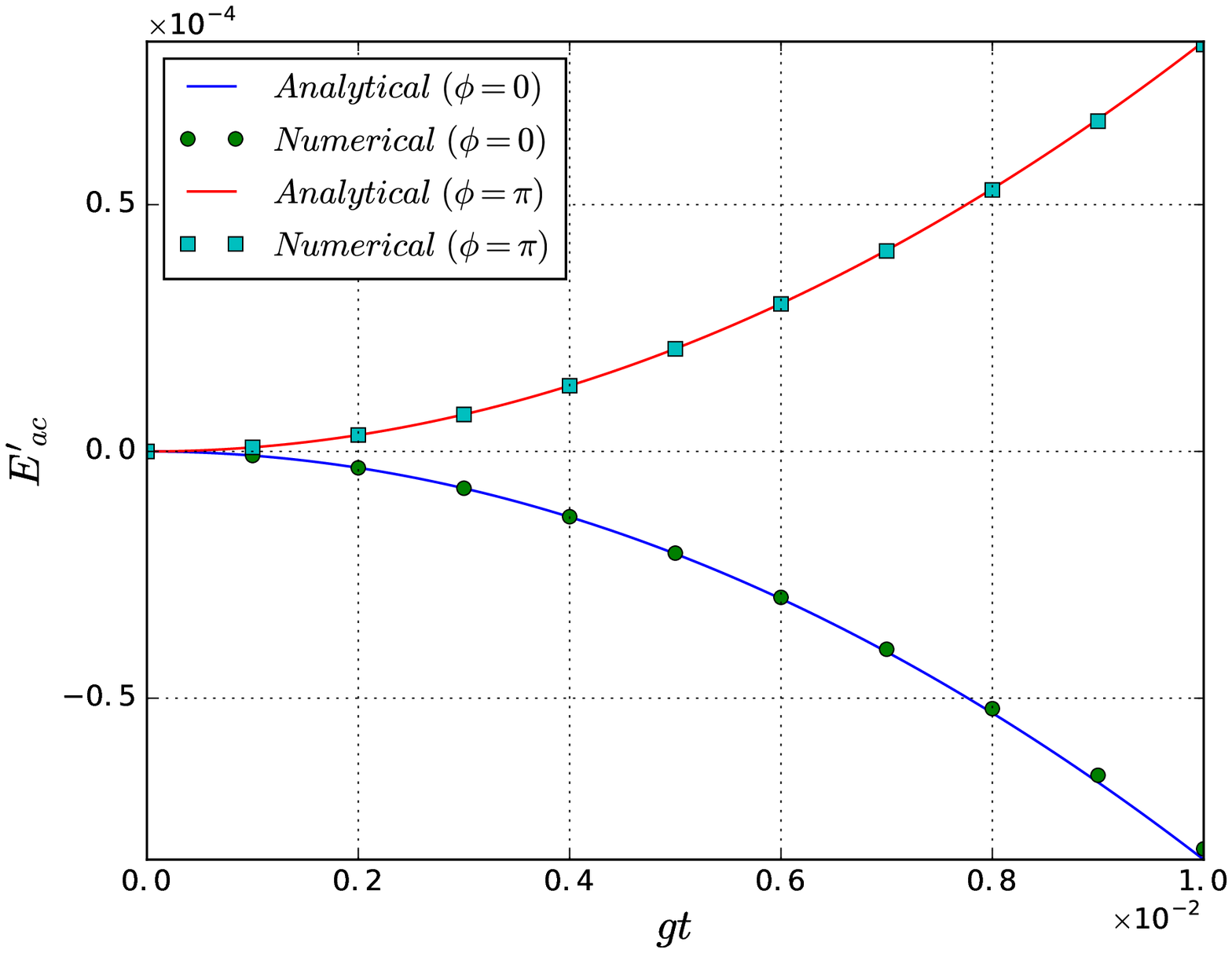}
\par\end{centering}
}

\caption{\label{fig:entanglement}(Color online) Evolution of entanglement
witnesses is shown with $gt$ using (a) HZ1 and (b) HZ2 criteria.
The values of parameters are same as used in the previous figures. }
\end{figure}

\subsection{Quantum Entanglement}

With the advent of quantum computing and quantum communication, a
large number of applications of entanglement have been reported (see
\cite{nielsen,my book} for review). We have mentioned about some
of them in Section \ref{sec:Introduction}. Motivated by these applications,
and because of the possibility that the exciton-photon entanglement
generated in the physical system of our interest can be of use in
the solid state quantum computing, it is of particular importance
to investigate the possibility of observing photon-exciton entanglement
in the light-semiconductor interaction. In order to obtain the signature
of the photon-exciton two-mode entanglement, we use three criteria.
Note that we have used here more than one inseparability criterion
as each of these criteria is only sufficient witness of entanglement
and not necessary. Two criteria used here were proposed by the Hillery
and Zubairy \cite{HZ-PRL,HZ2007,HZ2010} (from here on, we will refer
to these criteria as HZ1 and HZ2, respectively), and the third one
was proposed by Duan et al., \cite{duan} (which will be referred
to as Duan et al.'s criterion). 

In our case, HZ1 and HZ2 criteria can be expressed as \cite{HZ-PRL,HZ2007,HZ2010}
\begin{equation}
E_{a,c}=\left\langle N_{a}(t)N_{c}(t)\right\rangle -\left|\left\langle a(t)c^{\dagger}(t)\right\rangle \right|^{2}<0\label{24}
\end{equation}
and 
\begin{equation}
E_{ac}^{\prime}=\left\langle N_{a}(t)\right\rangle \left\langle N_{c}(t)\right\rangle -\left|\left\langle a(t)b(t)\right\rangle \right|^{2}<0,\label{25}
\end{equation}
respectively. Using these two criteria and the solutions obtained
above in Eq. (\ref{3}), we obtain the analytical expressions for
the entanglement witnessing parameter $E_{a,c}$ and $E_{a,c}^{\prime}$
as follows

\begin{equation}
\begin{array}{lcl}
E_{a,c} & = & \left|f_{3}\right|^{2}\left|\alpha\right|^{4}\left|\beta\right|^{2}+\left[\left(f_{1}^{*}f_{6}-f_{1}f_{3}^{*}h_{1}^{*}h_{2}\right)\mid\alpha\mid^{2}\alpha\beta^{*}\right.\\
 & + & \left.{\rm c.c.}\right]
\end{array}\label{26}
\end{equation}
and 
\begin{eqnarray}
E_{a,c}^{\prime} & = & \left|f_{3}\right|^{2}\left|\alpha\right|^{4}\left|\beta\right|^{2}-\left(h_{1}^{*}h_{4}\mid\alpha\mid^{2}\alpha\beta^{*}+{\rm c.c.}\right),\label{27}
\end{eqnarray}
respectively. As these two criteria are only sufficient not necessary,
we also investigate the existence of entanglement using Duan et al.'s
criterion \cite{duan}, which can be expressed as 
\begin{equation}
\begin{array}{lcl}
d_{ij}=\left\langle \left(\triangle u_{ij}\right)^{2}\right\rangle +\left\langle \left(\triangle v_{ij}\right)^{2}\right\rangle -2 & < & 0\end{array},\label{28}
\end{equation}
where 
\begin{equation}
\begin{array}{lcl}
u_{ij} & = & \frac{1}{\sqrt{2}}\left\{ \left(i+i^{\dagger}\right)+\left(j+j^{\dagger}\right)\right\} ,\\
v_{ij} & = & -\frac{i}{\sqrt{2}}\left\{ \left(i-i^{\dagger}\right)+\left(j-j^{\dagger}\right)\right\} .
\end{array}\label{29}
\end{equation}
Duan et al.'s criteria and the solution (\ref{3}) obtained above
can now be used together to obtain 
\begin{equation}
\begin{array}{lcl}
d_{ac} & = & 2\left[\left|f_{3}\right|^{2}\left|\alpha\right|^{4}+\frac{1}{2}\left\{ f_{1}^{*}h_{2}+h_{1}f_{2}^{*}\right.\right.\\
 & + & \left(f_{1}^{*}f_{5}+2f_{1}f_{6}^{*}\right)\alpha^{*}\beta+\left(2f_{1}h_{4}^{*}+h_{1}f_{5}^{*}\right.\\
 & + & \left.\left.2f_{3}h_{2}^{*}\right)\left|\alpha\right|^{2}+{\rm c.c.}\right].
\end{array}\label{30}
\end{equation}
To illustrate the presence of photon-exciton entanglement in our system,
we plot the right-hand sides of Eqs. (\ref{26}), (\ref{27}), and
(\ref{30}). Clearly, the signature of the intermodal entanglement
is seen in the photon-exciton coupled mode using only HZ1 and HZ2
criteria (shown in Fig. \ref{fig:entanglement}), but not through
Duan et al.'s criterion. Fig. \ref{fig:entanglement} further establishes
the effect of the phase of the input coherent light on the presence
of nonclassicality. Specifically, the presence of entanglement is
reflected through HZ1 (HZ2) criterion for the phase angle $\phi=\pi$
($0$), while it failed to show its presence for $\phi=0$ ($\pi$).
It further establishes that the set of criteria studied here is sufficient
only. Further justifying this fact, Duan et al.'s criterion failed
to detect entanglement (not shown here). In brief, the use of HZ1
and HZ2 criteria studied here establishes that the photon-exciton
mode is entangled for both $\phi=0$ and $\pi$.

\section{Nonclassicality in open quantum system \label{sec:nonclassicality-under-cw} }

In this section, we aim to study the nonclassical properties associated
with the bosonic modes when the system described above is considered
to be under the influence of the environment. Specifically, we now
consider a scenario, where both the semiconductor and photon modes
are allowed to interact with the ambient environment. In order to
solve the Hamiltonian (\ref{1}) under this situation, we need to
construct a master equation in the Lindblad form \cite{Perina-Book}.
This master equation, under the framework of rotating wave approximation,
takes the following form in the interaction picture (in the units
of $\hbar=1$)
\begin{equation}
\begin{array}{lcl}
\frac{d\rho}{dt} & = & -i\left[H,\rho\right]+\underset{j=1,2,3}{\sum}\left(L_{j}\rho L_{j}^{\dagger}-\frac{1}{2}L_{j}^{\dagger}L_{j}\rho-\frac{1}{2}\rho L_{j}^{\dagger}L_{j}\right),\end{array}\label{32}
\end{equation}
where $L_{1}=\left\{ \left(n_{th}+1\right)\gamma\right\} ^{\frac{1}{2}}a,$
$L_{2}=\left(n_{th}\gamma\right)^{\frac{1}{2}}a^{\dagger},$ and $L_{3}=\sqrt{\gamma}c$
are the Lindblad operators, and $\gamma$ denotes the rate of dissipation
for both exciton and photon modes. Here, we have assumed same rate
of dissipation for both optical and exciton modes. Further, we have
considered a vacuum bath ($n_{th}=0$) for the optical mode and thermal
bath for the exciton mode. The parameter $n_{th}$ corresponds to
the mean number of quanta in the heat bath corresponding to the exciton
mode. The numerical simulations are performed by the Quantum toolbox
QuTiP 3.1.0 \citep{qutip1,qutip2}. Using the numerical solution with
the help of subtool \emph{qutip.mesolve}, we have investigated the
nonclassical properties of the aforementioned system.

\subsection{Nonclassicality in open quantum system: Squeezing and antibunching}

We have reported single mode squeezing and antibunching in the exciton
mode and intermodal squeezing and antibunching in exciton-photon mode
in Section \ref{sec:Nonclassical-effects}. However, we failed to
report both single mode squeezing and antibunching in the photon mode
using the perturbative analytic solution. In this section, we have
used the criteria of nonclassicality defined in Eqs. (\ref{12}),
(\ref{15}), (\ref{18}), and (\ref{22}) to study time evolution
of these witnesses of nonclassicality. The obtained results are illustrated
in Fig. \ref{antibunchingmono}. Specifically, Fig. \ref{antibunchingmono}
(a) shows the variation of quadrature variance $\left(\left(\Delta X_{a}\right)^{2}\right)$
in the exciton mode in the pure state case (considering dissipation
rate zero) and non-zero dissipation rate. It can be clearly observed
that with increase in the value of $\gamma$, the maximum amount of
squeezing represented by the least value of the variance decreases
which can be seen by decay in the envelope in the corresponding figure.
Further, in Fig. \ref{antibunchingmono} (b), we have shown the evolution
of variance of quadrature variable $Y_{c}$ for photon mode, where
nonclassicality can be observed to decay with the increase in dissipation.
Further, we can clearly observe that the exact numerical result even
in the pure state case do not show squeezing in photon mode for $gt<0.3$,
i.e., in the domain of the validity of the perturbative solution reported
in Section \ref{sec:The-Model-Physical}. Thus, the obtained results
remain consistent with the analytical results reported previously.
However, as the numerical result is valid for much larger timescale,
we can now observe the presence of squeezing in photon mode as well.
Note that we have not shown the variation of variance in the other
quadrature here which were observed to show squeezing.

\begin{widetext}

\begin{figure}[H]
\subfloat[]{\includegraphics[scale=0.45]{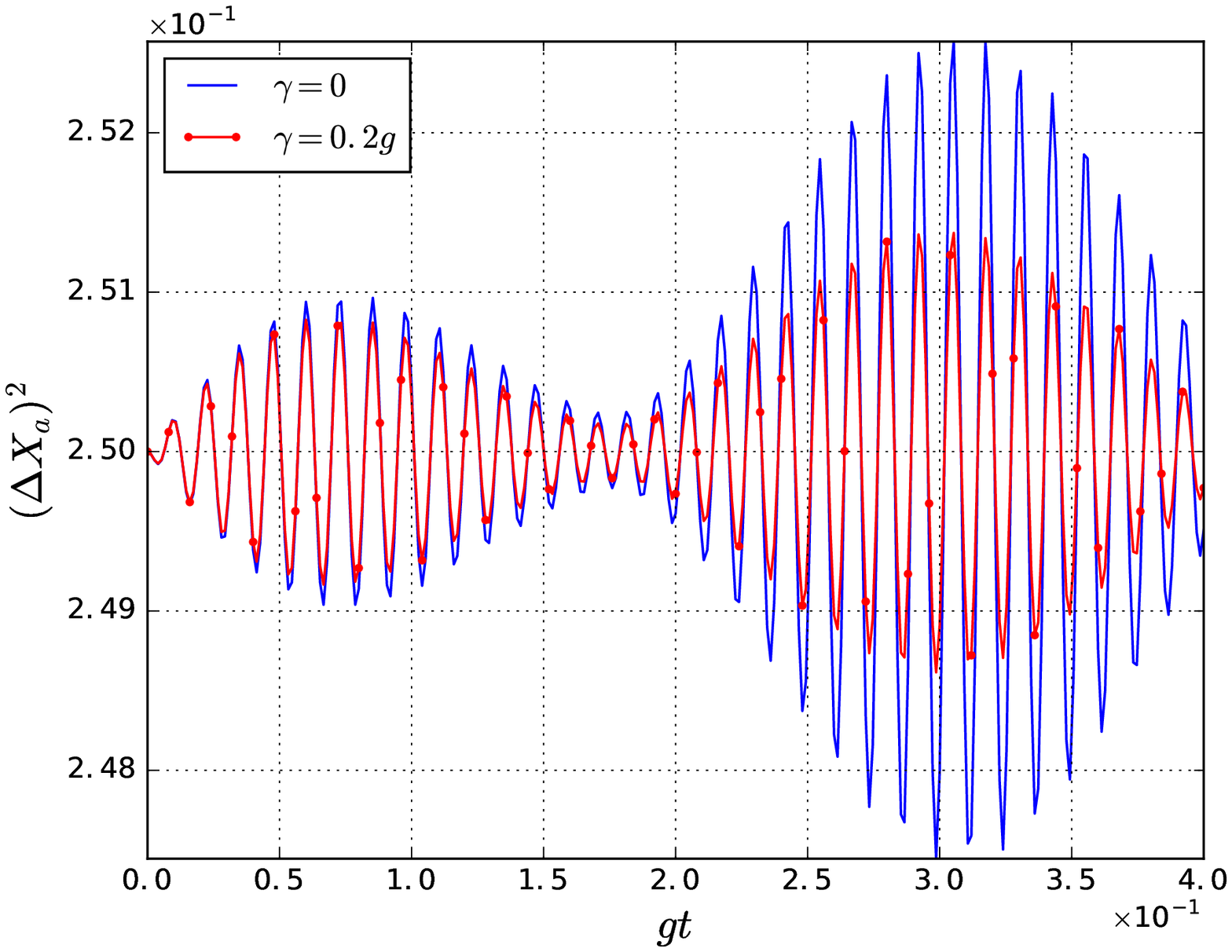}

} \subfloat[]{\includegraphics[scale=0.45]{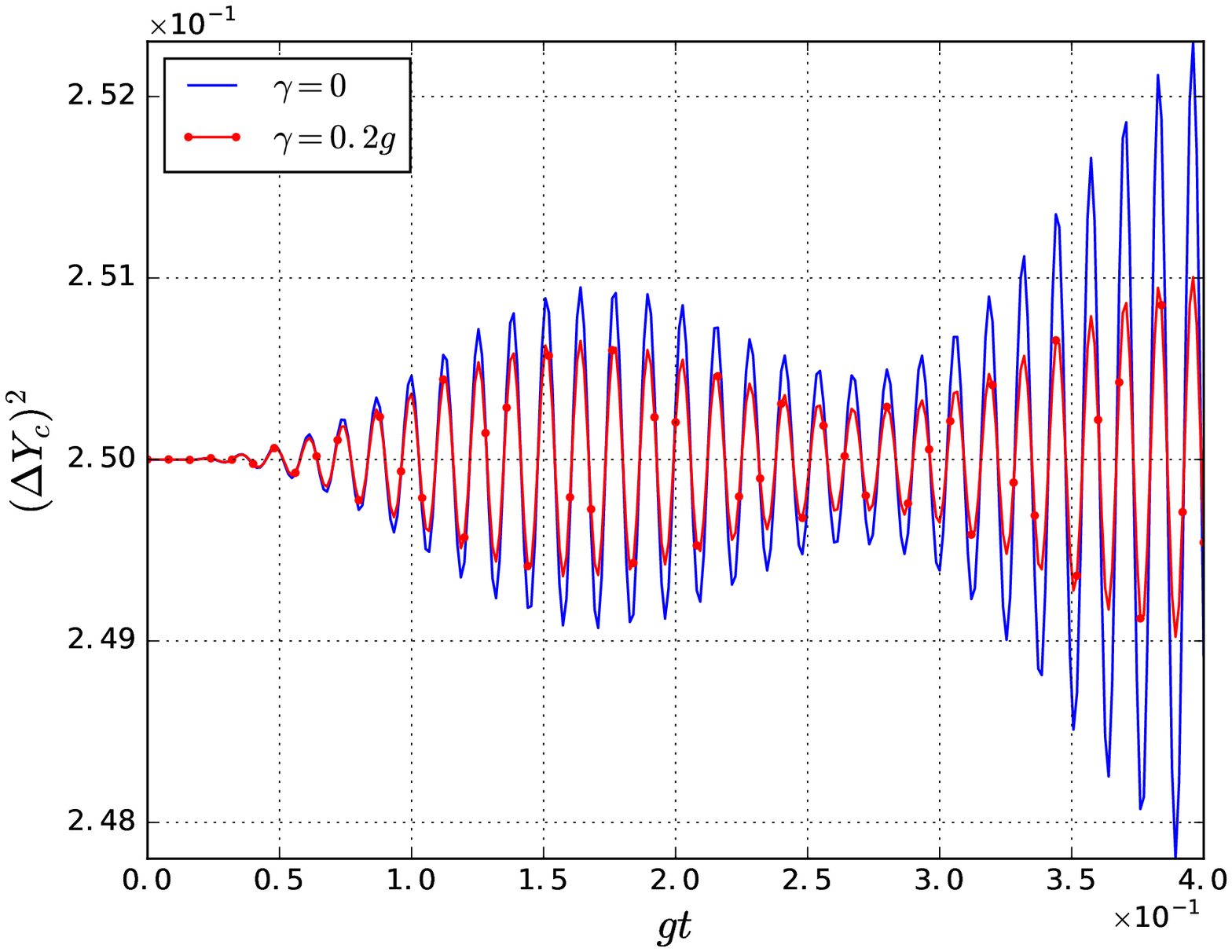}

}

\subfloat[]{\includegraphics[scale=0.45]{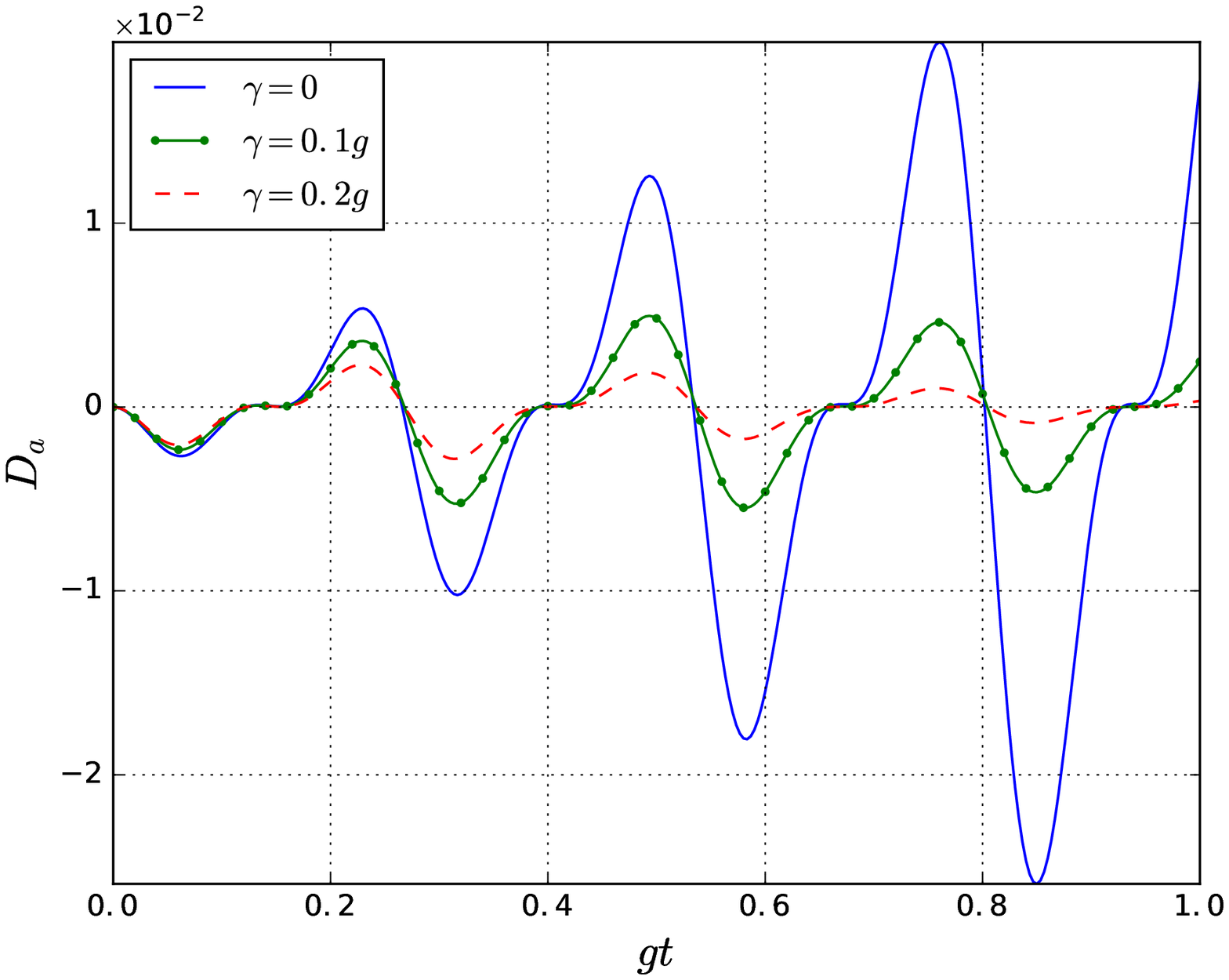}

} \subfloat[]{\includegraphics[scale=0.45]{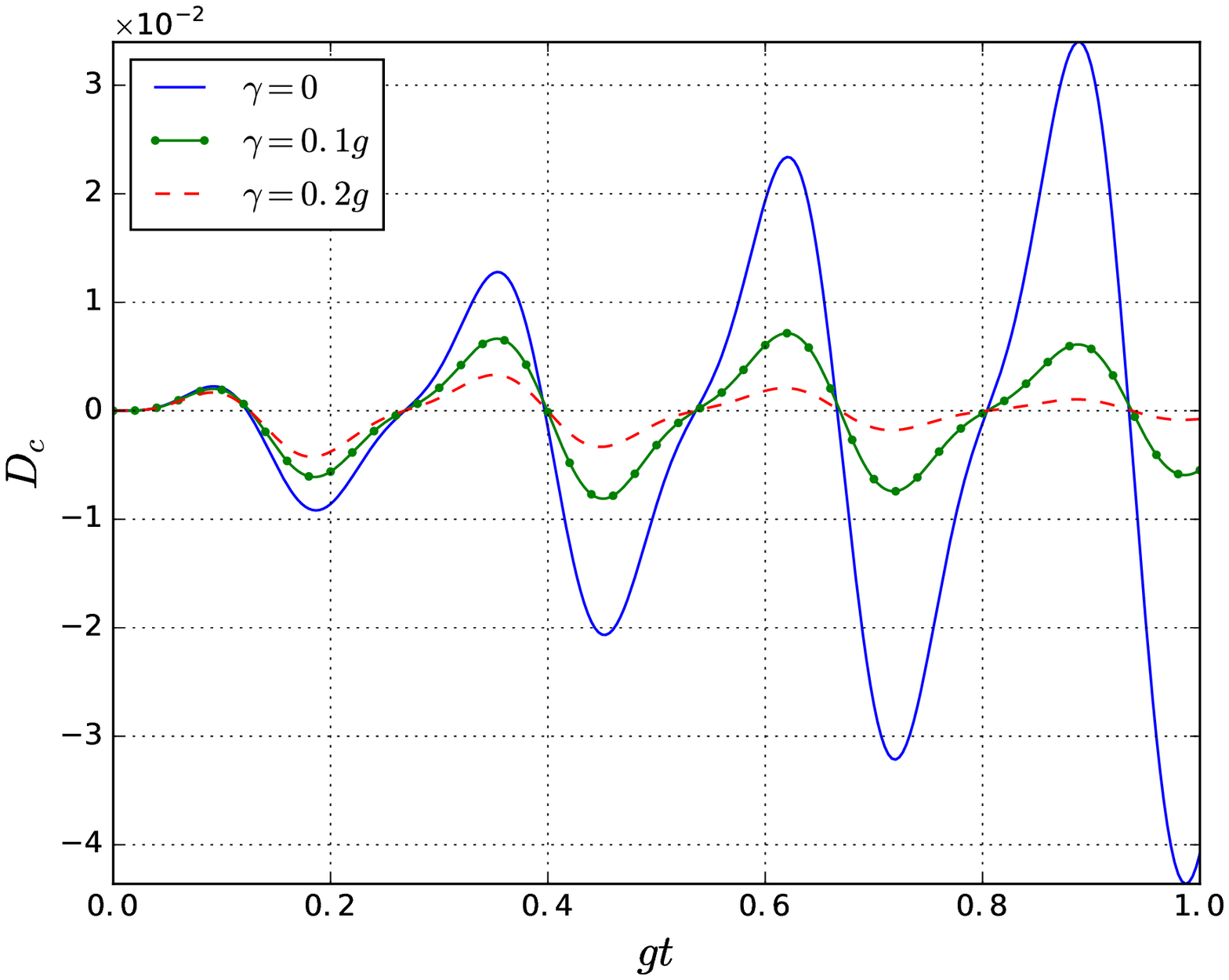}

}

\subfloat[]{\includegraphics[scale=0.45]{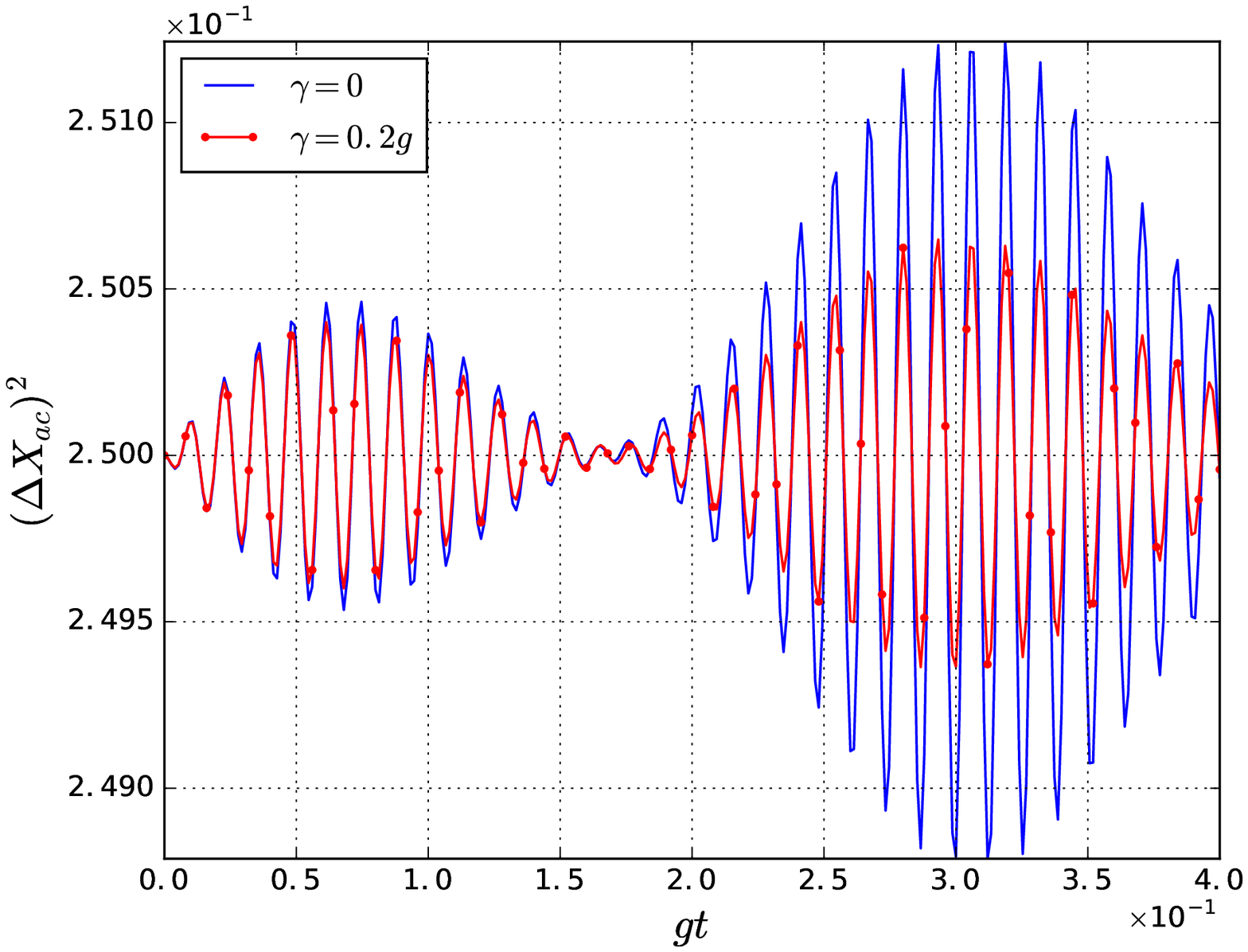}

} \subfloat[]{\includegraphics[scale=0.45]{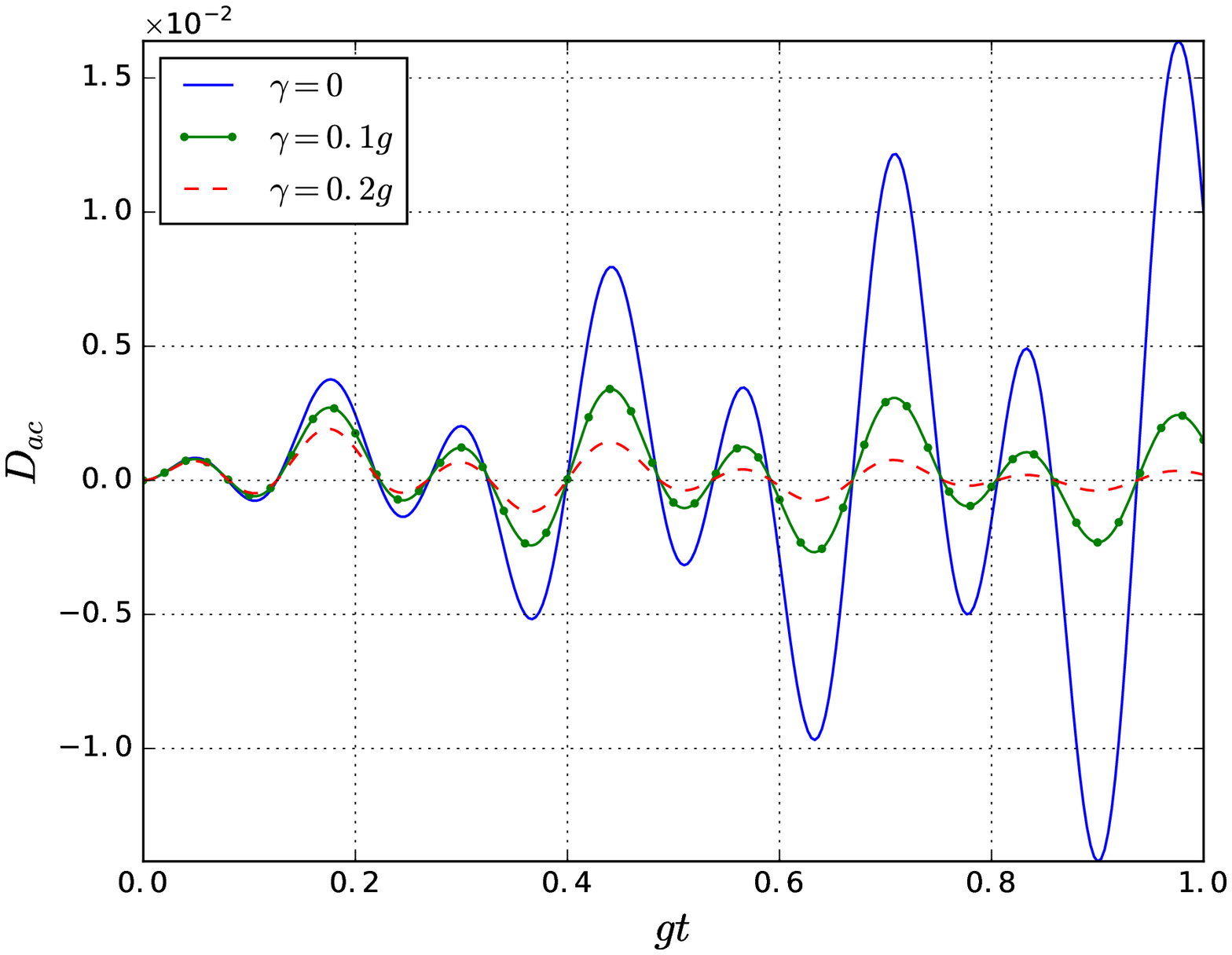}

}

\caption{\label{antibunchingmono}(Color online) Variation of the variance
in the measured values of the quarature variable (a) $X_{a}$ for
exciton mode and (b) $Y_{b}$ for photon mode. The values less than
$\frac{1}{4}$ are signature of squeezing. Time evolution of single
mode antibunching in (c) exciton and (d) photon modes. Intermodal
(e) squeezing and (f) antibunching in the exciton-photon mode are
also shown here. In all these cases, we have assumed vacuum bath for
the exciton mode and the values of the dissipation rate are mentioned
in the plotlegends.}
\end{figure}

\begin{figure}
\subfloat[]{\includegraphics[scale=0.45]{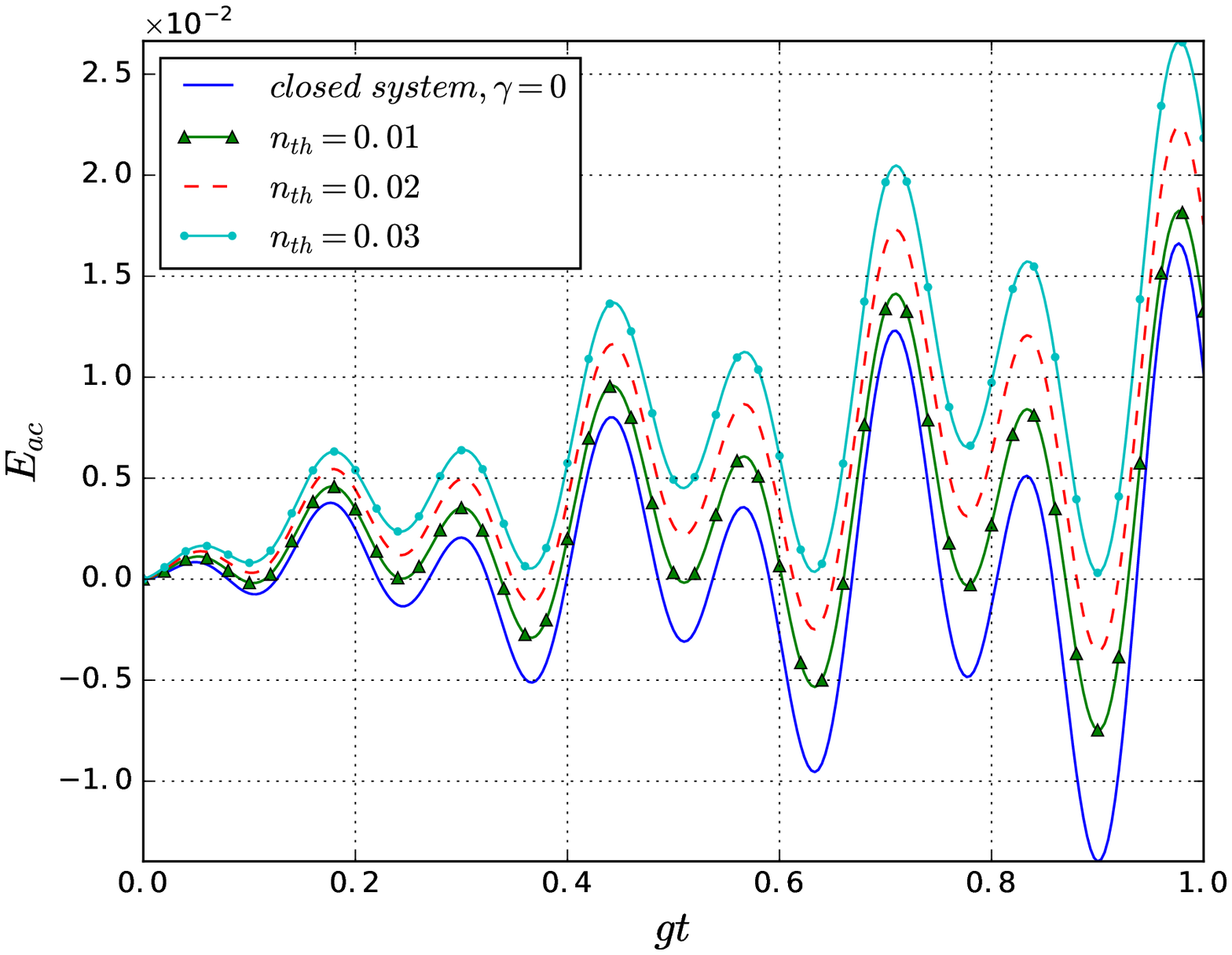}

} \subfloat[]{\includegraphics[scale=0.45]{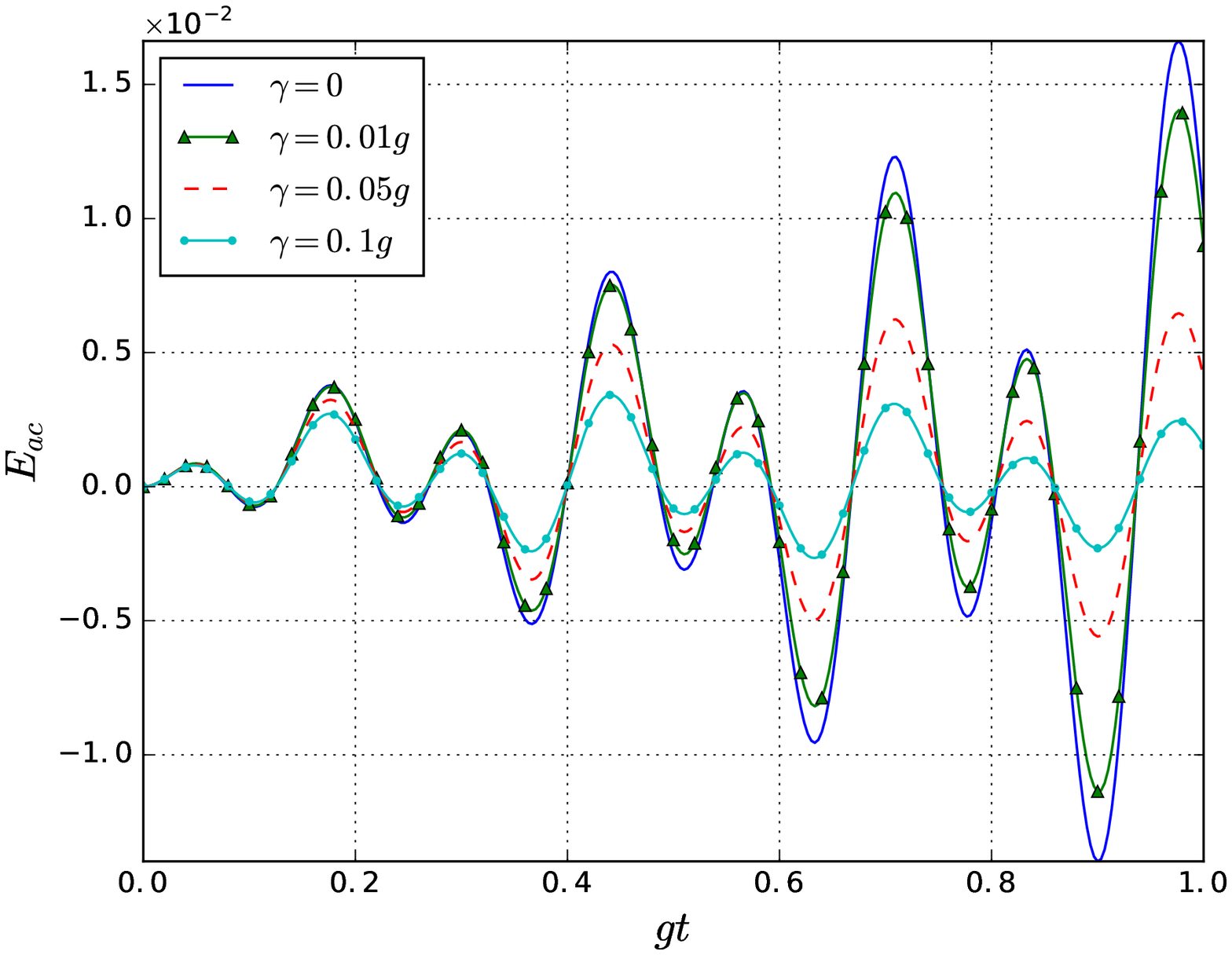}

}

\subfloat[]{\includegraphics[scale=0.45]{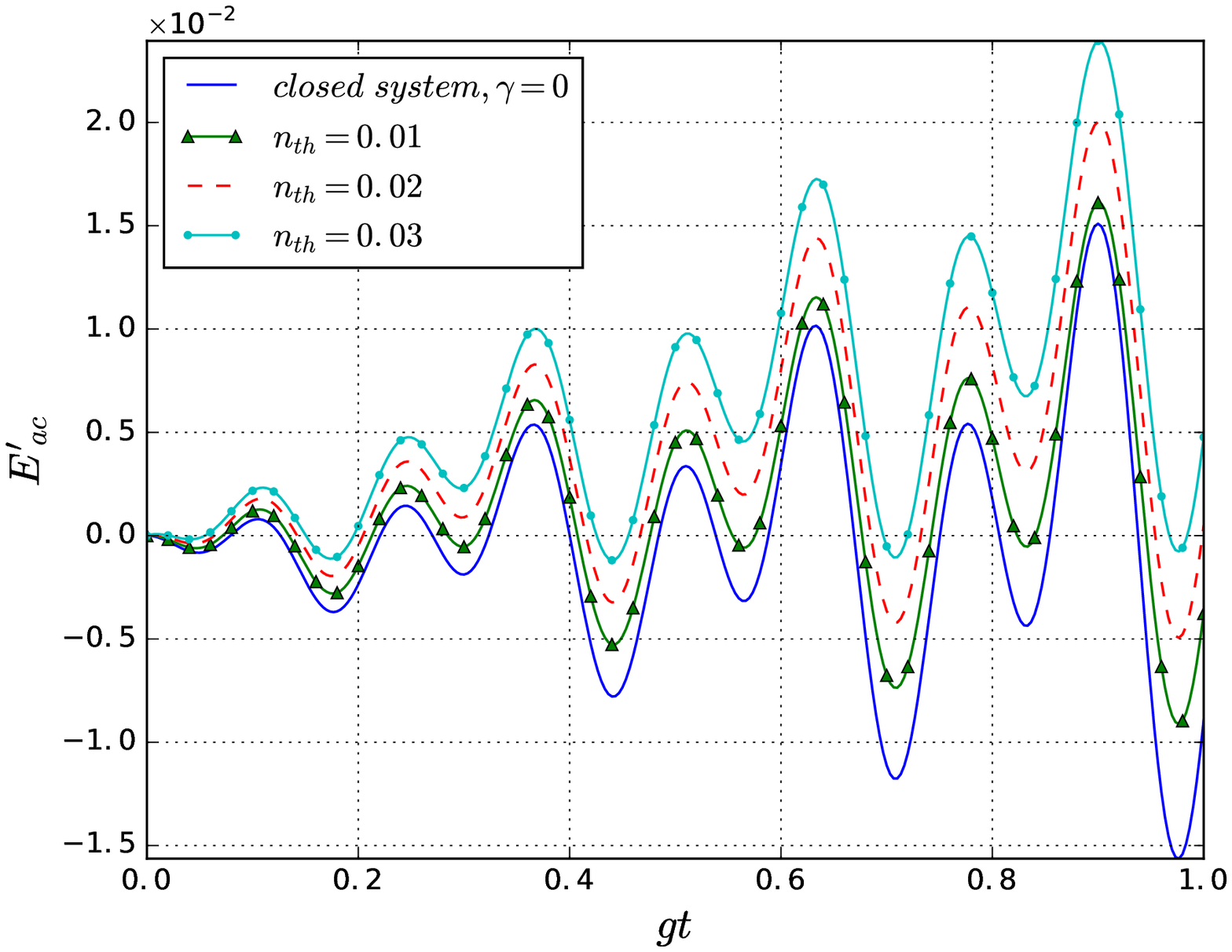}

} \subfloat[]{\includegraphics[scale=0.45]{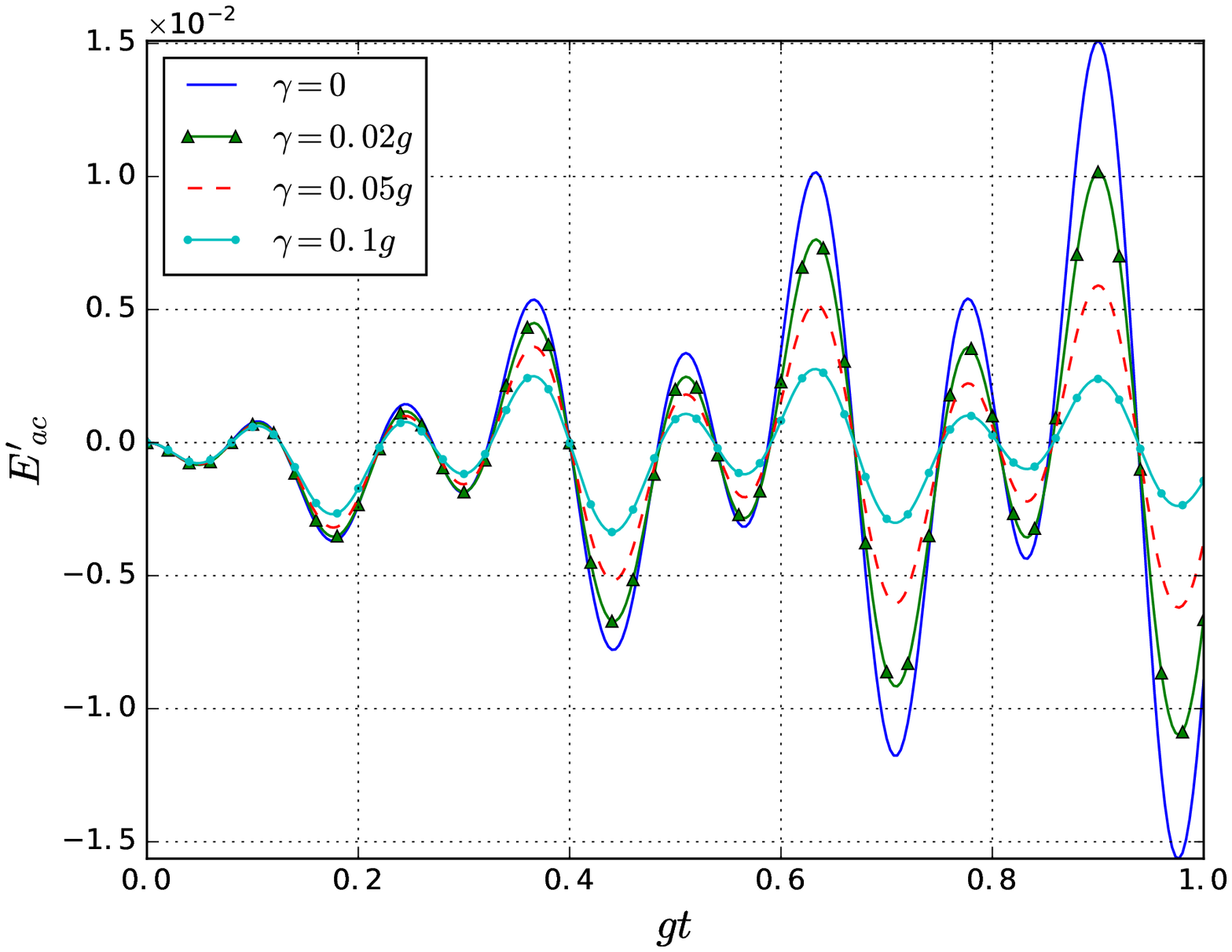}

}

\caption{\label{entmono}(Color online) The effect of thermal excitations ((a)
and (c)) and dissipation rate ((b) and (d)) on two witnesses of entanglement,
namely HZ1 ((a)-(b)) and HZ2 ((c)-(d)). Different values of the parameters
are mentioned in the plots. In ((a) and (c)), we have chosen $\gamma=0.01$,
while vacuum bath is considered in ((b) and (d)).}
\end{figure}

\end{widetext}

Thereafter, we report single mode antibunching in the exciton and
photon modes in Figs. \ref{antibunchingmono} (c) and (d), respectively.
In both closed and open quantum system scenarios, antibunching in
the exciton mode over long interaction time $gt$ shows an oscillatory
behavior, unlike a monotonically decreasing nature observed in Fig.
\ref{figantibunching} (a) for short timescale. Further, Fig. \ref{antibunchingmono}
(d) fails to show antibunching in the small timescale in the photon
mode. This is consistent with the results reported in Section \ref{sec:Nonclassical-effects}.
However, antibunching can be observed in large timescale for optical
mode. Finally, for the sake of completeness of the discussion, we
have also reported intermodal squeezing and antibunching in Figs.
\ref{antibunchingmono} (e) and (f), respectively. The presence of
intermodal squeezing in the longer timescale, and its absence in the
short timescale, is also consistent with the analytical results reported
in previous section. Finally, all these observed nonclassical features
are found to decay with the increase in the dissipation rate assuming
a vacuum bath for the exciton mode. 

\subsection{Entanglement in open quantum system}

Here, we will study the effect of change in the average thermal photon
number of the thermal bath on the exciton-photon entanglement and
compare it with that of the different values of dissipation rate.
Using the solution of Eq. (\ref{32}) with the Hillery-Zubairy criteria
of entanglement (\ref{24})-(\ref{25}), we have shown the variation
of the entanglement witness in Fig. \ref{entmono}. Specifically,
Fig. \ref{entmono} (a) shows that with increase in $n_{th}$ the
witness of HZ1 criterion drifts towards the positive side, thus signifying
the decay in entanglement and overall quantum to classical transition.
Note that, the entanglement may still be present in the timescale
we have shown, which was observed to decay eventually for larger values
of rescaled time. However, our interest here is to show the effect
of different parameter on the observed nonclassicality so we have
shown our results in a small timescale. Further, in case of vacuum
bath for exciton mode, increase in the value of dissipation rate is
found to cause a reduction in the value of the witness of nonclassicality
(HZ1 entanglement criterion) shown in Fig. \ref{entmono} (b). Thus,
a collective effect of thermal bath and higher values of dissipation
rate ensure nonclassicality to decay more rapidly. Similar studies
performed for HZ2 criterion of entanglement and illustrated in Fig.
\ref{entmono} (c)-(d) establish the same fact. Further, in the short
timescale of the analytical solution, we have observed entanglement
using HZ1 criterion only for some specific values of the phase angle of
the input coherent beam. However, in the larger timescale, entanglement
in all the cases is quite evident.

\subsection{Steering in open quantum system}

Schr{\"o}dinger introduced the notion of steering in response to
EPR paradox \citep{Sch-En}. He argued that the choice of measurement
on the first subsystem can change the final state of the second subsystem,
now known as EPR-steering. This correlation is stronger than entanglement
and weaker than Bell nonlocality. Incidentally, in the pure state
case the presence of entanglement is sufficient to establish stronger
correlation, i.e., EPR steering \citep{pure-ste}. Therefore, we did
not discuss the feasibility of observing steering independently in
the previous section. 

\begin{figure}
\includegraphics[scale=0.47]{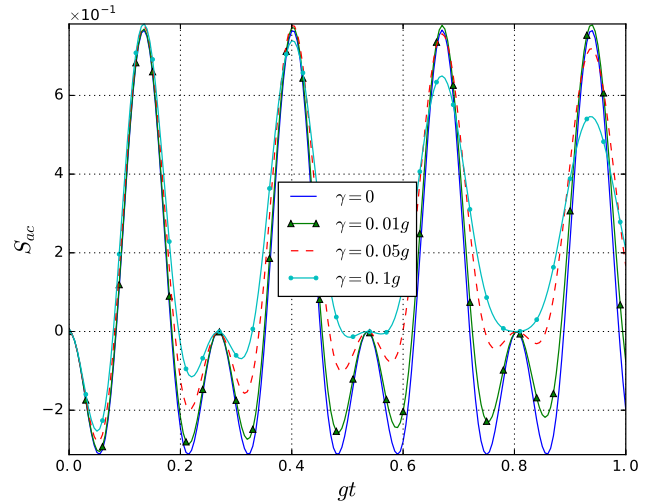}
\caption{\label{steer}(Color online) The effect of change in the dissipation
rate on witness of EPR steering. Different values of the parameters
are mentioned in the plots considering interaction of excitons with
the vacuum bath. In this case, we have assumed the exciton mode initially
in vacuum and photon mode in Fock $\left|n=5\right\rangle $ state.}
\end{figure}

The EPR-steering criteria can be introduced in terms of the HZ1 criterion
(\ref{24}) mentioned in the previous section (\citep{He-st,javid}
and references therein) as 
\begin{equation}
S_{ac}=E_{ac}+\frac{\langle a^{\dagger}a\rangle}{2}<0.\label{Steering}
\end{equation}
Steering is an asymmetric form of correlation as mode $a$ can steer
mode $c$ more than the amount by which mode $c$ can steer $a$.
The negative values of the witness $S_{ac}$ in Eq. (\ref{Steering})
would establish that mode $a$ can steer mode $c$, while the other
possibility is obtained by using the number operator of $c$ mode
in Eq. (\ref{Steering}). In the present case, we observed that the
exciton mode can steer the quantum state of the photon mode (shown
in Fig. \ref{steer}). Clearly, the negative values of the witness,
signifying EPR steering, decrease with the dissipation rate when both
modes are interacting with the vacuum bath. Specifically, the initial
state in this particular case is different from rest of the cases,
where both modes were initially coherent. Here, we have assumed the
exciton mode is initially in vacuum state and photon mode is in five
photon Fock state. In view of the fact that we failed to observe EPR
steering with initial coherent states, the presence of this strong
correlation in this case manifests the advantage of using nonclassical
input states. Particularly, Fock states are known as most nonclassical
states and as antibunched state, therefore the photon mode is initially
in single mode nonclassical state while exciton is in vacuum. The
interaction leads to strong correlations between two modes showing
EPR steering. It is not our purpose here to discuss the advantages
of nonclassicality in the initial states but this shows that the single
mode nonclassicality generated in the system itself has applications
in generation of strong bipartite correlations. 

\section{Conclusion \label{sec:Conclusion}}

Nonclassical properties associated with the light-semiconductor interaction
are rigorously studied using two physical scenarios. Firstly, the
interaction of a coherent light with a two-band semiconductor is investigated,
and the existence of several nonclassical features, such as squeezing,
intermodal squeezing, antibunching, intermodal antibunching, and entanglement,
is established. Interestingly, antibunching and squeezing is observed
in the exciton mode using the perturbative solution, but not in photon
mode. However, intermodal antibunching, squeezing and entanglement
were observed. In some cases (viz., intermodal antibunching and entanglement),
the nonclassicality in the output can be controlled by the phase of
the input field. This systematic study is performed using analytical
technique and is found to coincide exactly with the corresponding
numerical results obtained using QuTip 3.1.0. This establishes the
validity of the present perturbative solution in the domain in which
it is applied here. Finally, in the last section, we have extended
the work to a more general scenario, where the system is considered
to be under the effect of the ambient environment. This is our second
physical scenario and this particular scenario is investigated using
numerical methods only. Interestingly, all the signatures of nonclassicality
observed using the perturbative solution are found to be present in
this situation, too. On top of that, we have also observed squeezing
and antibunching in the photon mode in this case. It is also established
that the photon mode can be steered by the exciton mode as correlation
between these two modes are stronger than quantum entanglement and
show EPR steering. Particularly, we have observed squeezing and antibunching
in both exciton and photon modes, intermodal squeezing, antibunching,
entanglement, and EPR steering in exciton-photon mode, but what is
more interesting and worth mentioning here is that the observed nonclassical
features are robust to noise. During this effort we have established
the advantage of nonclassical states if used as the input field in
the present case as it leads to the EPR steering correlation between two
modes. In fact, the origin of the nonclassical features observed here
(except EPR steering) can be attributed to the nonlinear interaction
between exciton-exciton pairs represented by the coupling constant
$\chi$. 

The results obtained here seem to be interesting for two particular
reasons. Firstly, the system studied here is very general as it can
be reduced to various physical systems of particular interest, and
thus the obtained results can be used to investigate the possibilities
of observing nonclassical features in many other systems. Specifically,
the Hamiltonian (\ref{1}) studied here also corresponds to the single-mode
resonant field propagating through a Kerr-like medium \cite{Agarwal Puri}.
The Hamiltonian (\ref{1}) also reduces to the interaction between
two modes of a beam splitter if $\chi=0$, which is equivalent to
the optical couplers studied in Ref. \citep{Obrien} and references
therein. Further, for $g=0$, it reduces to the Hamiltonian of a system
composed of a harmonic oscillator (free field part) and a quartic
anharmonic oscillators (which represents light interacting with a
Kerr-type medium) that are not interacting with each other. Secondly,
it is interesting because of its potential applications in quantum
information processing. Specifically, it is expected that the observed
entanglement between a semiconductor and the light interacting with
it, would be of use to the quantum information community interested in
solid state quantum computing which is extremely exciting at the moment
because it leads to a possibility of room temperature quantum computing.
Further, the present study has revealed EPR steering between the exciton
and photon mode, which has its own several applications in device
independence. Moreover, the generated squeezed state may be used to
build a solid state source of squeezed state for use in continuous
variable quantum cryptography (see \citep{CV-rev} and references
therein).

\textbf{Acknowledgment:} AP thanks Department of Science and Technology
(DST), India for the support provided through the project number EMR/2015/000393.
KT thanks the project LO1305 of the Ministry of Education, Youth and
Sports of the Czech Republic for support.

\end{document}